\documentclass[pra,twocolumn,aps,superscriptaddress,nofootinbib]{revtex4-1}
\usepackage{amsfonts}
\usepackage{mathtools}
\usepackage[svgnames,table]{xcolor}
\usepackage[T1]{fontenc}
\usepackage{txfonts}
\usepackage{graphicx}
\usepackage{lineno}

\usepackage{hyperref}
\hypersetup{
        colorlinks=true,
        citecolor=SteelBlue,
        filecolor=LimeGreen,
        linkcolor=SlateBlue,
        urlcolor=MediumPurple
}

\usepackage{dcolumn}% Align table columns on decimal point
\usepackage{bm}% bold math
\usepackage{amsmath}
\usepackage{breqn}
\usepackage[bbgreekl]{mathbbol}
\usepackage{bbm}
\usepackage{mathrsfs}
\usepackage[permil]{overpic}

\newcommand{\equ}[1]{\begin{equation}#1\end{equation}}
\newcommand{\comment}[2]{\textcolor{orange}{}}

\makeatletter
\let\cat@comma@active\@empty
\makeatother

\begin{document}
\preprint{APS/123-QED}
\title{Enhancing interferometer sensitivity without 
sacrificing bandwidth and stability: beyond single-mode
and resolved-sideband approximation}
\author{Xiang Li}
\affiliation{Burke Institute for Theoretical Physics, California Institute of Technology, Pasadena, California}
\author{Jiri Smetana}
\affiliation{Institute for Gravitational Wave Astronomy, School of Physics and Astronomy, University of Birmingham, Birmingham B15 2TT, United Kingdom}
\author{Amit Singh Ubhi}
\affiliation{Institute for Gravitational Wave Astronomy, School of Physics and Astronomy, University of Birmingham, Birmingham B15 2TT, United Kingdom}
\author{Joe Bentley}
\affiliation{Institute for Gravitational Wave Astronomy, School of Physics and Astronomy, University of Birmingham, Birmingham B15 2TT, United Kingdom}
\author{Yanbei Chen}
\affiliation{Burke Institute for Theoretical Physics, California Institute of Technology, Pasadena, California}
\author{Yiqiu Ma}
\affiliation{Center for Gravitational Experiment, School of Physics, Huazhong University of Science and Technology, Wuhan, China}
\author{Haixing Miao}
\affiliation{Institute for Gravitational Wave Astronomy, School of Physics and Astronomy, University of Birmingham, Birmingham B15 2TT, United Kingdom}
\author{Denis Martynov}
\affiliation{Institute for Gravitational Wave Astronomy, School of Physics and Astronomy, University of Birmingham, Birmingham B15 2TT, United Kingdom}
\date{\today}

\begin{abstract}
Quantum noise limits the sensitivity of precision measurement devices, such as laser interferometer gravitational-wave observatories and axion detectors. %and optomechanical setups. 
In the shot-noise-limited regime, these resonant detectors are subject to a trade-off between the peak sensitivity and bandwidth. One approach to circumvent this limitation in gravitational-wave detectors is to embed an anomalous-dispersion optomechanical filter to broaden the bandwidth. The original filter cavity design, however, makes the entire system unstable. Recently, we proposed the coherent feedback between the arm cavity and the optomechanical filter to eliminate the instability via PT-symmetry\,\cite{Li2020Broadband}. The original analysis based upon the Hamiltonian formalism adopted the single-mode and resolved-sideband approximations. In this paper, we go beyond these approximations and consider realistic parameters. We show that the main conclusion concerning stability remains intact, with both Nyquist analysis %incorporating time delay 
and a detailed time-domain simulation. 
\end{abstract}

\maketitle

\iffalse
{\color{red}To-dos:
\begin{itemize}
    \item (Everyone) Double-check that your original meaning remains faithfully after Xiang's polishing;
    \item (Everyone) Go through the comments in orange, polish the text for whichever you can;
    \item (Everyone) Add necessary citations in your responsible parts;
    \item [$\checkmark$](Xiang) Add filter cavity response to further illustrate the necessity of optical spring compensation [done]; modify Fig.5 for that purpose [done];
    \item [$\checkmark$](Jiri) Change font style in the plots to be Times New Roman for consistency throughout the text [done];
    \item [$\checkmark$](Amit) Polish the content on page 10, refer to the comments there for details;
\end{itemize}
}
\fi

\section{\label{sec:intro}Introduction}

The detection of gravitational waves (GW) from a binary black hole merger in 2015\,\cite{Abbott2016Observation} opened a new window of astronomy observation. Binary black hole systems have so far been the most commonly measured GW sources\,\cite{LVC2019GWTC1,LVC2020GWTC2}. The demand for extracting richer properties of the ringdown stage\,\cite{Kamaretsos2012Black,Kamaretsos2012Is,Taracchini2014Small,Abbott2016First,Hughes2019Learning}, as well as other astrophysical processes that produce pronounced gravitational waves at high frequencies, e.g., the binary neutron star  mergers\,\cite{LVC2017GW170817, Miao_kHz_2018, Martynov_kHz_2019} and core collapse supernovae\,\cite{Ott2009The,Kotake2013Multiple,Radice2019Characterizing}, calls for the broadband and high-frequency sensitivity of gravitational wave detectors. 
For current advanced detectors and even future detectors including Einstein Telescope\,\cite{Punturo2010Einstein,Maggiore2020Science} and Cosmic Explorer\,\cite{Reitze2019cosmic}, the quantum shot noise limits the detector sensitivity from a few hundred Hz to kilo Hz\,\cite{Abbott2017Exploring}. 
Similarly, recently proposed detectors of axion-like-particles in the galactic halo suffer from the photon shot noise across their sensitivity bands~\cite{DeRocco_axions_2018, Obata_axions_2018, Liu_axions_2019, Martynov_axions_2020}.

In the canonical interferometer configuration\,\cite{Kimble2001Conversion}, resonant arm
cavities are used to increase the relative signal strength by 
effectively extending the length of the detector via repeated 
reflections of the optical field. However, the positive dispersion of the arm cavity makes the signal at higher frequencies no longer resonant. This leads to an inverse relationship between the peak sensitivity and bandwidth of the detector, known as the Mizuno limit\,\cite{mizuno1995comparison}. This can be traced back to the energetic quantum limit\,\cite{braginsky1995quantum,braginsky2000energetic}, which is also called the quantum Cramer-Rao Bound (QCRB)\,\cite{Tsang2011Fundamental}, and is therefore limited by the quantum fluctuation of the intracavity light field\,\cite{Miao2017Towards}. 

One approach to broaden the bandwidth without sacrificing peak sensitivity is to attach a negative-dispersion optomechanical filter cavity to the arm cavity\,\cite{miao2015enhancing}, which can compensate for the phase gained in the arm cavity and thus resulting in a \emph{white light cavity} effect. However, such a scheme is dynamically unstable and thus an additional stabilizing controller must be implemented. We will call this unstable white light cavity scheme as \emph{uWLC} for short. In the original proposal\,\cite{miao2015enhancing}, in addition to the filter cavity, there are several auxiliary optics, either for impedance match with the input mirror of the arm cavity or for steering the field to the filter, which leads to a rather complex setup. 

\begin{figure}[t!]
\includegraphics[width=0.8\columnwidth]{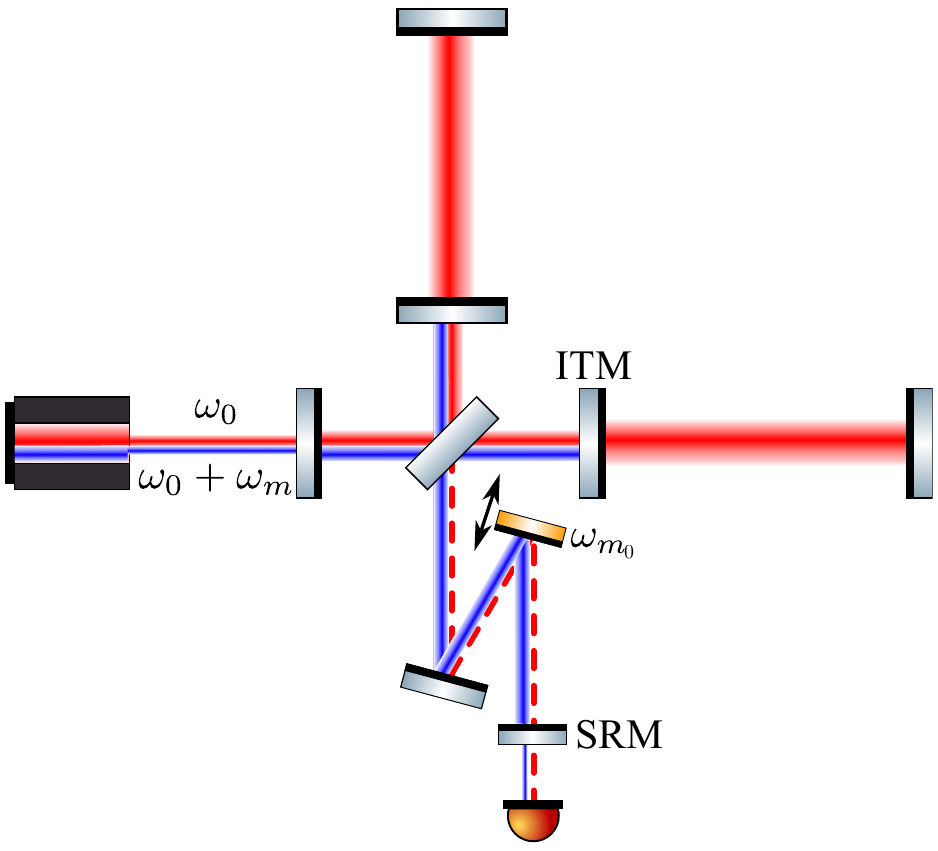}
\caption{A schematics of the interferometer configuration with the signal-recycling cavity (SRC) as the optomechanical filter. Both the main arm cavities and the SRC resonant at $\omega_0$, but they have different free spectral ranges due to the different cavity lengths. One of the mirrors in the SRC acts as the mechanical degree of freedom that resonant at $\omega_{m_0}$. It couples to the signal sidebands around $\omega_0$ via the radiation pressure due to the blue detuned SRC pumping field at $\omega_0+\omega_{m}$. Note that $\omega_{m_0}/\omega_{m}\approx 1$, with slight difference caused by the optical spring effect due to the blue-detuned pumping (will be explained later near Eq.\,\eqref{eq:omega_opt}). ITM: input test mass, SRM: signal-recycling mirror.} 
\label{fig:config}
\end{figure}

In a later study\,\cite{Bentley2019Converting}, it was found that converting the signal-recycling cavity (SRC) into the optomechanical filter can lead to bandwidth broadening with a much simpler optical layout, as illustrated in Fig.\,\ref{fig:config}. The parameter regime considered\,\cite{Bentley2019Converting}, however, still leads to an unstable system. We recently realised that, when the optomechanical interaction strength is smaller than or equal to the coupling frequency between the arm cavity and the filter cavity, the system will be self-stablised\,\cite{Li2020Broadband}. More interestingly, the peak sensitivity is improved together with the bandwidth, not due to squeezing but a significant enhancement of the signal response. We shall call this stably operated white light cavity scheme as \emph{sWLC} for short.

So far, the sWLC scheme has only been analysed using the Hamiltonian in the single-mode and the resolved-sideband approximation\,\cite{Bentley2019Converting,Li2020Broadband}, which treats the arm cavity signal field, the mechanical oscillator, and the field in the filter cavity as single modes, separately. The stability issue of the system is based on the poles of the resulting input-output relation\,\cite{Li2020Broadband}. One natural question to ask is, whether the stability and sensitivity improvement remain valid when these approximations are removed by considering realistic parameters. Answering such a question defines the theme of this paper. 

The outline of this paper goes as follows: in Section\,\ref{sec:ideal_to_full} we revisit the idealised Hamiltonian dynamics and stability requirement, %which defines the parameter regime for the more detailed analysis. In Section\,\ref{sec:full}, and we present the complete description of the interferometer configuration without the 
and introduce the full analysis method beyond the single-mode and resolved-sideband approximations. In Section\,\ref{sec:freq}, we solve the system dynamics in the frequency domain. We also analyse the stability using the Nyquist criterion and show the resulting sensitivity. In Section\,\ref{sec:time}, we carry out a detailed time-domain simulation, and show the agreement with the frequency-domain analysis. Throughout the paper, we will be focusing on the shot-noise-limited sensitivity, leaving the back-action noise from radiation pressure for future studies.

\section{\label{sec:ideal_to_full}From single-mode approximation to full analysis}

In this section, we will first recap the idealised Hamiltonian dynamics under the single-mode approximation as analysed in Refs.\,\cite{Bentley2019Converting, Li2020Broadband} for the sWLC, and also revisit the stability requirement. %\comment{}{We will further analyse the decoherence effect caused by the thermal fluctuation of the mechanical mode, which plays a more stringent role compared with that of uWLC scheme.} 
We will then introduce the analysing framework considering the realistic setup, by abandoning the approximations applied in the previous treatment. The detailed analysis in the frequency and time domain will be carried in Section\,\ref{sec:freq} and \ref{sec:time}, respectively.

\begin{figure}[!t]
\includegraphics[width=\columnwidth]{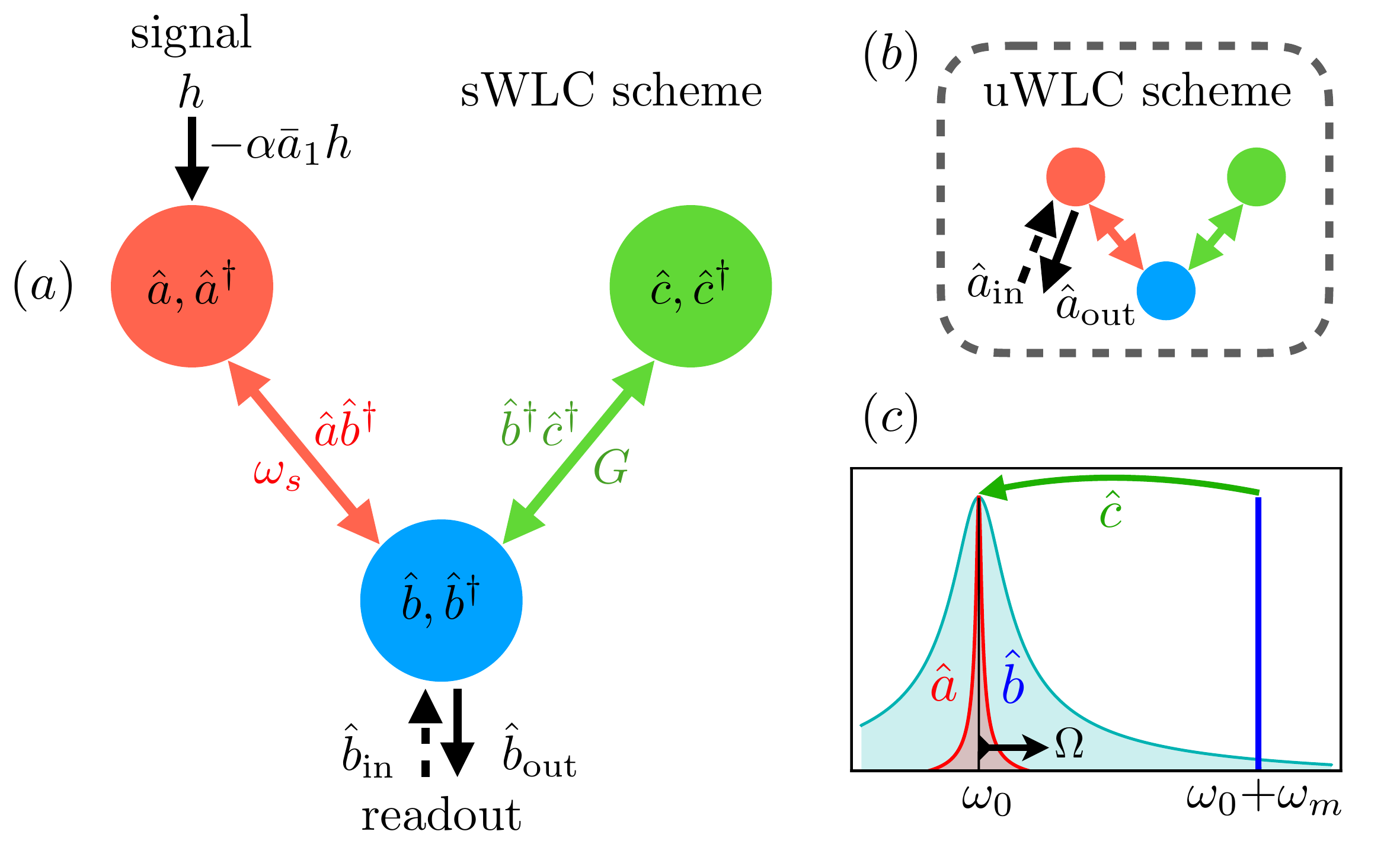}
\caption{
Idealised mode interaction structure of the optomechanical system illustrated in Fig.\,\ref{fig:config}. \textbf{(a)} The sWLC scheme. Here $\hat a$ represents the differential mode of the arm cavity which couples to the GW strain $h$, $\hat b$ represents the SRC (filter) mode, and $\hat c$ represents the mechanical mode of the suspended mirror. The coupling between $\hat a$ and $\hat b$ ($\hat b$ and $\hat c$) is characterised by $\omega_s$ ($G$). Mode $\hat b$ is coupled to the external ingoing field $\hat b_{\rm in}$, which carries the vacuum noise, and the outgoing field $\hat b_{\rm out}$, which carries the signal and  will be read out.  
\textbf{(b)} The uWLC scheme for comparison. In the intracavity readout scheme presented in Ref.\,\cite{miao2015enhancing}, the external fields couple to the arm cavity mode $\hat a$, rather than the filter cavity mode $\hat b$. \textbf{(c)} The frequency-domain mode structure under single-mode (for both $\hat a$ and $\hat b$) and resolved sideband (for $\hat b$) approximations. The parametric interaction $\hat b^{\dagger} \hat c^{\dagger}$ is realized by the optomechanical coupling under blue-detuned pumping by the mechanical resonant frequency. Note that all mode operators are defined in the rotating frame of frequency $\omega_0$, and $\Omega$ is the sideband frequency with respect to it.}
\label{fig:modeint}
\end{figure}

The idealised sWLC mode interaction in the rotating frame of frequency $\omega_0$, as illustrated in Fig.\,\ref{fig:modeint}, can be described by the following Hamiltonian:
\equ{\label{eq:Hint}
\hat H_{\rm int} =  i\, \hbar\omega_s (\hat a\, \hat b^\dagger -\hat a^\dagger \hat b) + i\, \hbar G (\hat b^\dagger \hat c^\dagger -\hat b\, \hat c)\,,}
with $\hat a\,,\hat b\,,\hat c\,$ being the quantum operators of the differential optical mode of the arm cavity, the mechanical mode, and SRC (i.e., the filter cavity) optical mode, respectively. Here $\omega_s$ is the beam-splitter-type interaction strength between mode $\hat a$ and $\hat b$, and $G$ describes the optomechanical interaction strength between mode $\hat b$ and mode $\hat c$.

\begin{figure}[!t]
\includegraphics[width=\columnwidth]{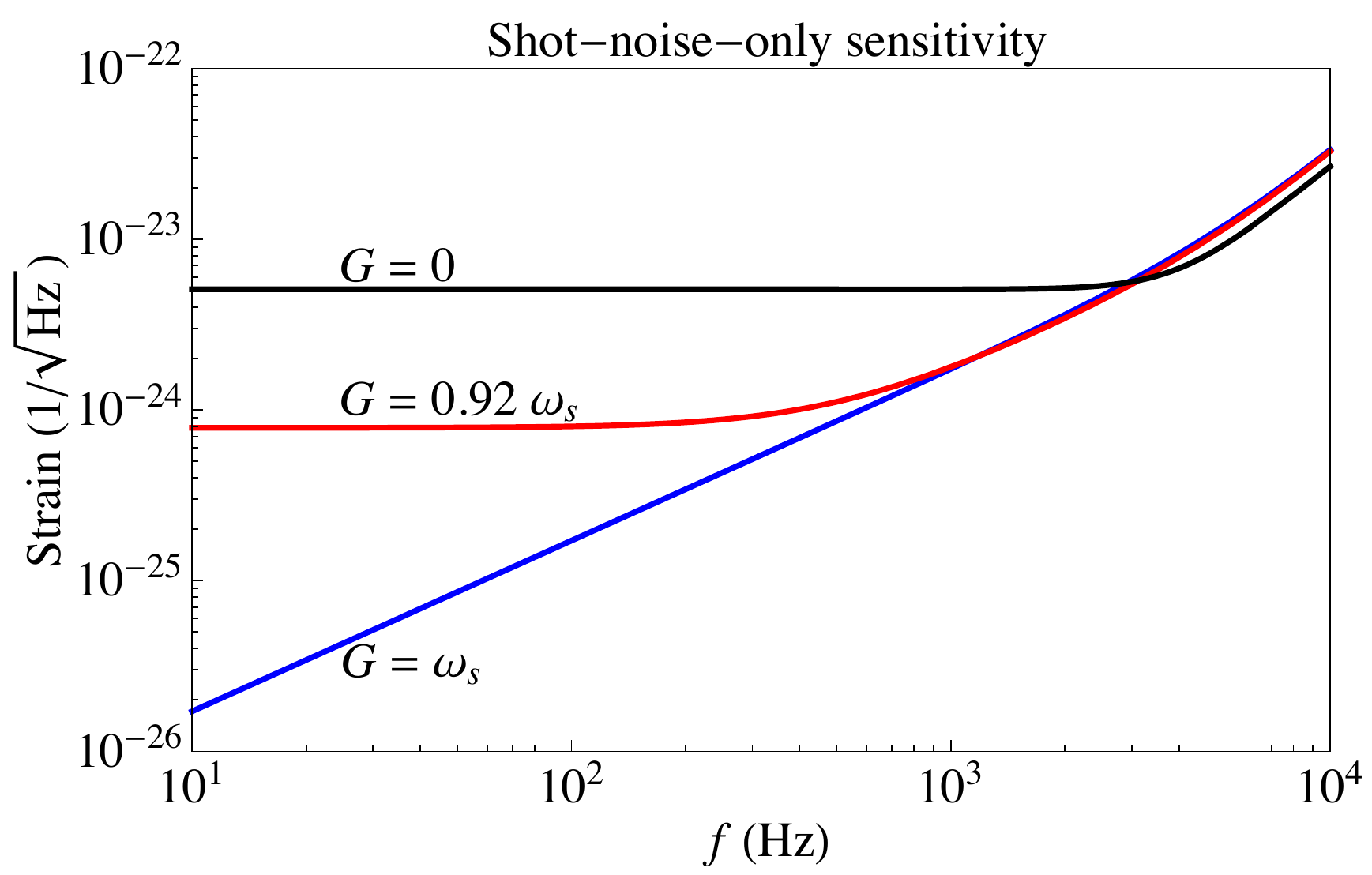}
\caption{The signal-referred shot noise spectral density for different optomechanical interaction strengths $G$. Different curves are plotted based on Eq.\,\eqref{eq:sen_sm} with the relation between $G$ and $\omega_s$ shown near each line. \comment{}{Add parameter used and conventional interferometer as reference; Delete the title on the top and moveit to the caption here}}
\label{fig:SNRidH}
\end{figure}

Considering the GW strain signal input $h$ with coupling strength $\alpha$, as well as the coupling to the external bath $\hat{b}_{\rm in}$ with rate $\gamma$\,\cite{Chen2013Macroscopic}, we obtain the Heisenberg equations of motion for the three modes:
\equ{\label{eq:abcEq}
\begin{aligned}
\dot{\hat{a}}(t) & =-\omega_s \hat{b}(t)+i \alpha h(t)\,, \\
\dot{\hat{c}}^{\dagger}(t) & = G\, \hat{b}(t)\,, \\
\dot{\hat{b}}(t) & =-\gamma \hat{b}(t)+\omega_s \hat{a}(t)+G \hat{c}^{\dagger}(t)+\sqrt{2 \gamma} \,\hat{b}_{\rm in}(t)\,,
\end{aligned}
}
with outgoing field given by $\hat b_{\rm out}(t)=-\hat b_{\rm in}(t)+\sqrt{2 \gamma} \,\hat b(t)$. The above equations can be solved in frequency domain via Fourier transform:
\begin{equation}\label{eq:FTaOmega}
\hat o(\Omega) ={\cal F}[\hat o (t)]\equiv \int_{-\infty}^{\infty}{\rm d}t\, e^{ i \Omega t} \hat o (t)\,,
\end{equation}
where $\Omega$ is the sideband frequency in the rotating frame of frequency $\omega_0$, and $\hat o$ represents $\hat a,\hat b,\hat c$ or $\hat b_{\rm in,out}$. The resulting solution for the outgoing field is:
\begin{align}
\nonumber
\hat b_{\rm out}(\Omega) = & \frac{i\Omega(\gamma + i \Omega) - G^2+\omega_s^2}{i\Omega(\gamma - i \Omega) + G^2 - \omega_s^2} \,\hat b_{\rm in}(\Omega) \\ & +\frac{i\sqrt{2\gamma}\, \omega_s \alpha  }{i\Omega(\gamma -i \Omega) + G^2-\omega_s^2}h(\Omega)\,. 
\label{eq:bout}
\end{align}
Interestingly, regardless of the value of $G$ and $\omega_s$, the outgoing field is not squeezed, as the modulus of input-to-output transfer function remains equal to unity. The signal is contained entirely in the phase quadrature\,\cite{caves1985New,Schumaker1985New} defined as $\hat Y= (\hat b_{\rm out} -\hat b_{\rm out}^{\dag})/(\sqrt{2}\,i)$. 
The resulting signal-referred shot noise spectral density when measuring the phase quadrature is given by 
\begin{equation}\label{eq:sen_sm}
S_{hh}(\Omega) = \frac{\Omega^2\gamma^2 + (G^2-\omega_s^2+\Omega^2)^2}{4\gamma \omega_s^2\alpha^2}\,. 
\end{equation}
The sensitivity given different relations between $G$ and $\omega_s$ is plotted in Fig.~\ref{fig:SNRidH}. One interesting case is when $G=\omega_s$ where the noise spectral density vanishes at DC as the signal response diverges:
\begin{equation}\label{eq:sen_pt}
S_{hh}(\Omega) |_{G=\omega_s} = \frac{\Omega^2(\Omega^2+\gamma^2)}{4\gamma\omega_s^2 \alpha^2}\,. 
\end{equation}
The system stability, as analysed in Ref.\,\cite{Li2020Broadband}, is determined by the poles of the transfer function in Eq.\,\eqref{eq:bout}, i.e. the roots of 
\begin{equation}\label{eq:denominator}
    i \Omega(\gamma -i \Omega) + G^2-\omega_s^2=0\,.
\end{equation}
When the imaginary part of any root becomes positive, the system becomes unstable. The locations of roots are determined by the relation between $G$ and $\omega_s$. It can be proven\,\cite{Li2020Broadband} that whenever $G$ is equal to or less than $\omega_s$, there is no unstable root. 
The critical point happens when $G=\omega_s$, and there is a pole at DC, which is consistent with the resulting signal response being infinite at DC, as can be seen in Eqs.\,\eqref{eq:bout} and\,\eqref{eq:sen_pt}. 
To summarise, under the idealised Hamiltonian, the system is stable with $G\leq \omega_s$ and at the same time, the shot-noise-limited sensitivity will be improved. In the following content, we will show that these features will remain intact even after we relax the approximations applied in deriving the idealised Hamiltonian.

%Before going beyond the single-mode approximation, let's consider the effects of thermal decoherence. As sWLC works near the divergence point, the effects of thermal decoherence from the mechanical oscillator is more serious compared with either conventional or uWLC scheme. \comment{Xiang}{add plots with different $T_{\rm env}$, as well as plots with different readout rate $\gamma$: inverse related compared with sWLC} This analysis indicates a more stringent requirement of mechanical quality factor and cryogenic condition. \comment{}{But are we only consider the infinite mechanical Q case here?}

In terms of the physical parameters described in Table\,\ref{tab:PhyPara}, the two approximations to produce the idealised Hamiltonian in Eq.\,\eqref{eq:Hint} are: (i) the single-mode approximations, i.e. $\Omega L_{\rm arm}/c\ll 1$ and $2\omega_{m_0} L_{\rm SRC}/c\ll 1$, which treats the arm cavity and the filter cavity as single modes each described by an annihilation operators $\hat a$ and $\hat b$, and (ii) the resolved-sideband approximation, i.e. $\omega_{m_0} L_{\rm SRC}/c\gg \gamma$, which treats the mechanical sidebands around $\omega_0$ as a single operator $\hat c$, ignoring the higher sidebands around $\omega_0+2\omega_{m_0}$ that should be involved in the interaction between the blue detuned filter cavity and the mechanical oscillator. 
Note that all the parameters $\omega_s,\,G,\,\alpha,\,\gamma$ used in Eqs.\,\eqref{eq:Hint}-\eqref{eq:denominator} above are effective parameters, that can be approximately expressed in terms of the physical parameters. Under the single-mode and resolved-sideband approximations, the mode interaction strength $\omega_s$ and $G$ can be expressed as:
\begin{equation}
\omega_s =
 \frac{c\, \sqrt{T_{\rm ITM}}}{2\sqrt{L_{\rm arm} L_{\rm SRC}}}, \quad G = \sqrt{\frac{8\pi P_b } {m\,\lambda \,\omega_{m_0} L_{\rm SRC} }}.
\label{eq:omegas_G}
\end{equation}
where $P_b$ is the power of the filter cavity pumping field that impinges on the mechanical oscillator. Also, the signal coupling strength $\alpha$ and the SRC cavity half-bandwidth (i.e. the decay rate, or bath coupling rate) $\gamma$ are defined as:
\begin{equation} \alpha = \sqrt{\frac{P_{\rm arm} L_{\rm arm} \omega_0}{ c \hbar}},\quad \gamma = \frac {c\, T_{\rm SRM}}{4 L_{\rm SRC}}\,.
\end{equation}

\begin{table}[!t]
\caption{A list of parameters and nominal values}
\begin{tabular}{ccc}
\hline
\hline
Parameters & Description & Value\\
\hline
$L_{\rm arm}$ & arm cavity length & 4\,km \\
\hline
$P_{\rm arm}$ & arm cavity power & 800\,kW \\
\hline
$T_{\rm ITM}$& ITM power transmissivity & 0.5\%\\
\hline
$L_{\rm SRC}$& SRC length & 40\,m\\
\hline
$T_{\rm SRM}$& SRM power transmissivity & 0.02\\
\hline
$P_b$ & filter cavity power &6.4\,kW\\
\hline
$\lambda $ & laser wavelength & 1064\,nm \\
\hline
$m$&oscillator mass & $10\,$mg \\
\hline
$\omega_{m_0}/(2\pi)$& mechanical frequency & $10^5$\,Hz\\
\hline
$Q_m$& mechanical quality factor & $\infty$ \footnote{We effectively remove the mechanical damping to highlight that the system can be self stabilised without additional damping mechanism. } \\
\hline
\hline
\end{tabular}
\label{tab:PhyPara}
\end{table}

For the full analysis, we adopt the approach in Ref.\,\cite{Kimble2001Conversion} by propagating the fields through the interferometer and taking into account their interactions with the mechanical degree of freedom via the radiation pressure. %Without loss of generality, 
As all optical elements are axisymmetric, for simplicity, the optical fields can be treated as 1D propagating ones along the optical axis. At each location, the optical field is represented as:
\begin{equation}\label{eq:omega0frame}
\hat E(t) =\hat o(t) e^{-i\omega_0 t} + \hat o^{\dag}(t) e^{i\omega_0 t}\,,
\end{equation}
where $\hat o(t)$ can represent $\hat a(t)$ and $\hat b(t)$, the slowly-varying field operators in the rotating frame of $\omega_0$ for the light inside the arm cavity and filter cavity respectively. We can define sideband operators $\hat a(\omega)$ and $\hat b(\omega)$ via Fourier transform as in Eq.\,\eqref{eq:FTaOmega}, but for a large sideband frequency $\omega$ up to the order of $\omega_{m_0}$. Note that so long as $\omega\ll \omega_0$, the slowly-varying operators in time domain and sideband operators in frequency domain will be well-defined. %$\hat a(t) =\int {\rm d}\omega/(2\pi) e^{-i\omega t} \hat a(\omega)$. 

\begin{figure}[!t]
\includegraphics[width=\columnwidth]{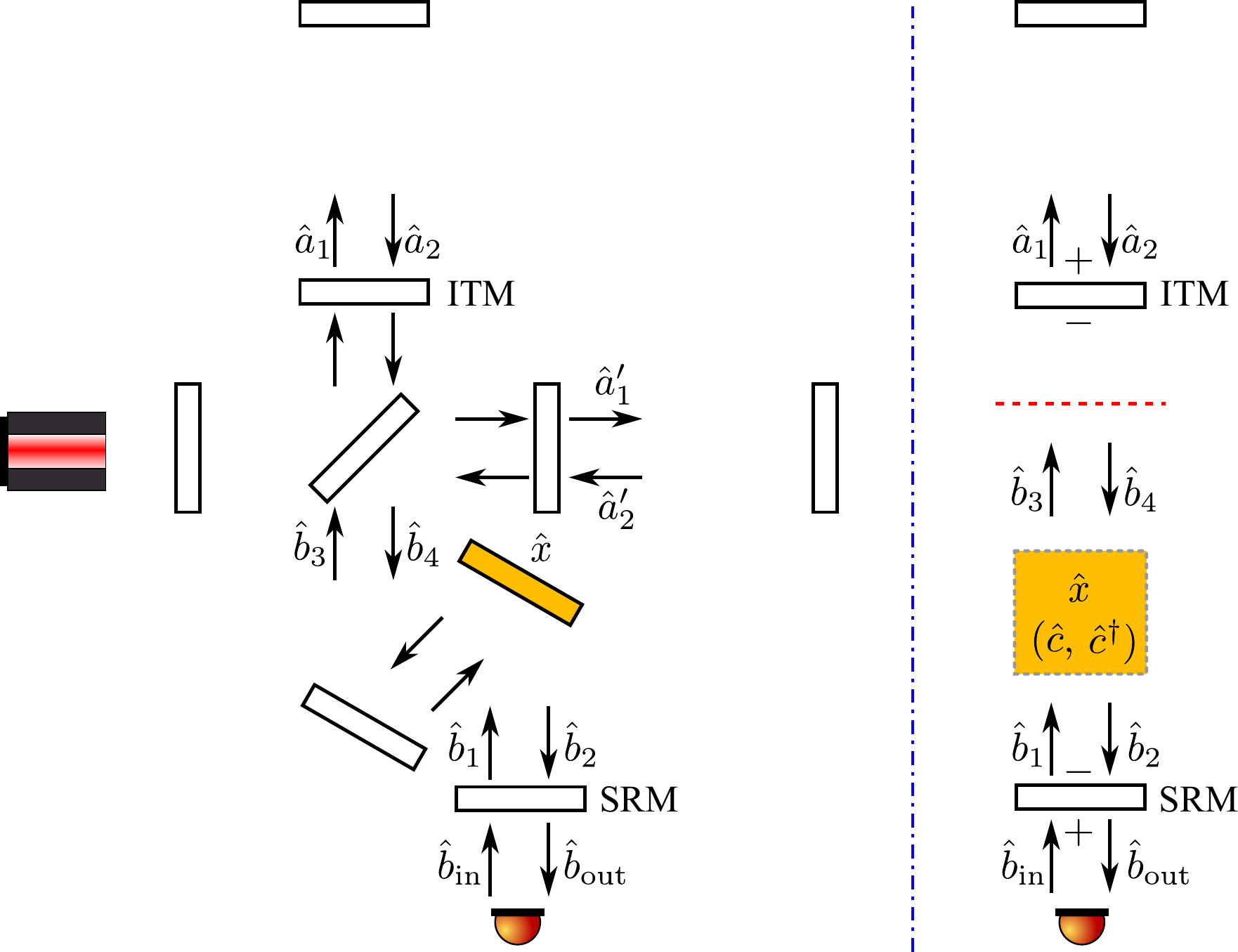}
\caption{The fields involved in the complete analysis of the scheme shown in Fig.\,\ref{fig:config}. The right figure is a simplified version of the left one when only looking at the differential mode of the two arms, which is adopted in the analysis in the main text. In the right figure, the sign convention for the mirror reflectivity is also shown. The arm cavity and the SRC length are both tuned to be integer numbers times the wavelength of the carrier at frequency $\omega_0$.} 
\label{fig:fields}
\end{figure}

The relevant fields fully describing the scheme in Fig.\,\ref{fig:config} are illustrated in Fig.\,\ref{fig:fields}. We will focus on the differential mode of the two arm cavities, and the right panel of Fig.\,\ref{fig:fields} shows the simplified representation considered here. We assume both the arm cavity and the SRC are tuned \comment{}{how?} such that the cavity lengths are integer numbers times the wavelength of the carrier at frequency $\omega_0$. Since we are looking at the linear dynamics, the equations of motion will only involve linear terms of the sideband operators.  %$\hat a(t)$ can be effectively viewed as the slowly-varying part of the classical electromagnetic field. 
The Heisenberg equation of motion is formally identical to the classical Maxwell equation. The field operators are described by the following set of equations:
\begin{subequations}
\begin{align}
\hat a_1(t) & = \sqrt{T_{\rm ITM}} \,\hat b_3(t) + 
\sqrt{R_{\rm ITM}} \,\hat a_2(t) \,, \label{eq:field_ops_1}\\
\hat a_2(t) & = \hat a_1( t - \tau_{\rm arm})+ 2 i k_0 A_{\rm arm} L_{\rm arm} h(t)\,, \\
\hat b_4(t) & = \sqrt{T_{\rm ITM}} \,\hat a_2(t) - \sqrt{R_{\rm ITM}} 
\,\hat b_3(t)\,, \\ 
\hat b_2(t) & = \hat b_4(t-\tau_{\rm SRC}/2) + 2 i k_b A_b e^{-i\omega_{m_0} t}\hat x(t)\,,\\
\hat b_3(t) & = \hat b_1(t-\tau_{\rm SRC}/2) + 2 i k_b A_b e^{-i\omega_{m_0} t}\hat x(t)\,, \\
\hat b_1(t) & = \sqrt{T_{\rm SRM}} 
\,\hat b_{\rm in}(t)-\sqrt{R_{\rm SRM}} \,\hat b_2(t) \,, \\
\hat b_{\rm out}(t) & = \sqrt{T_{\rm SRM}} 
\,\hat b_2(t)+ \sqrt{R_{\rm SRM}} \,\hat b_{\rm in}(t) \,. \label{eq:field_ops_2}
\end{align}
\end{subequations}
Here the round-trip delay times are defined as $\tau_{\rm arm}= 2 L_{\rm arm}/c$ and $\tau_{\rm SRC} = 2 L_{\rm SRC}/c$, for the arm cavity and SRC respectively. The two cavities are pumped with frequency $\omega_0$ and $\omega_0+\omega_{m_0}$, with wave vectors being $k_0 = \omega_0/c$ and $k_b = (\omega_0 +\omega_{m_0})/c\approx k_0$ ($\omega_0\gg \omega_{m_0}$) respectively. The steady-state field amplitudes $A_{\rm arm}$ and $A_b$ are given by $A_{\rm arm} = \sqrt{P_{\rm arm}/(2\hbar \omega_0)}$ and $A_b =  \sqrt{P_b/2\hbar \omega_0}$. The mechanical motion is driven by 
the radiation pressure in the presence of the blue-detuned pump field:
\begin{equation}
\ddot {\hat x}(t)+ \gamma_m \dot{\hat x}(t)+ \omega_{m_0}^2 \hat x(t) 
=\frac{\hat F_{\rm rad}(t)}{m}\,, \label{eq:oscillator_eom}
\end{equation}
where the radiation pressure $\hat F_{\rm rad}$ reads:
\begin{equation}
\hat F_{\rm rad}(t) =\frac{2  \hbar \omega_0  A^*_c}{c} e^{i\omega_{m_0} t} \left [\hat b_1(t) + \hat b_4(t) \right] + {\rm h.c.}\,
\end{equation}
%with $\rm h.c.$ denoting Hermitian conjugate.
The displacement operator $\hat x$ is related to the mechanical mode operator $\hat c$ by:
\begin{equation}
\hat x(t)= \sqrt{\frac{\hbar}{2m\omega_{m_0}}} \left[ \hat c(t) e^{-i\omega_{m_0} t} +  \hat c^{\dag}(t) e^{i\omega_{m_0} t} \right]\,. 
\end{equation}
%The mechanical damping rate $\gamma_m$ is related to the mechanical quality factor $Q_m$ by $\gamma_m = \omega_{m_0}/\gamma_m$. \comment{}{are we going to use them?}

\section{Frequency-domain analysis}
\label{sec:freq}

\begin{figure}[!t]
%\includegraphics[width=0.95\columnwidth]{Figures/mode_4d.pdf}

%\hspace{0.5pt}

%\includegraphics[width=0.8\columnwidth]{Figures/freq.pdf}
%\includegraphics[width=0.95\columnwidth]{Figures/optical_spring.pdf}
\includegraphics[width=0.95\columnwidth]{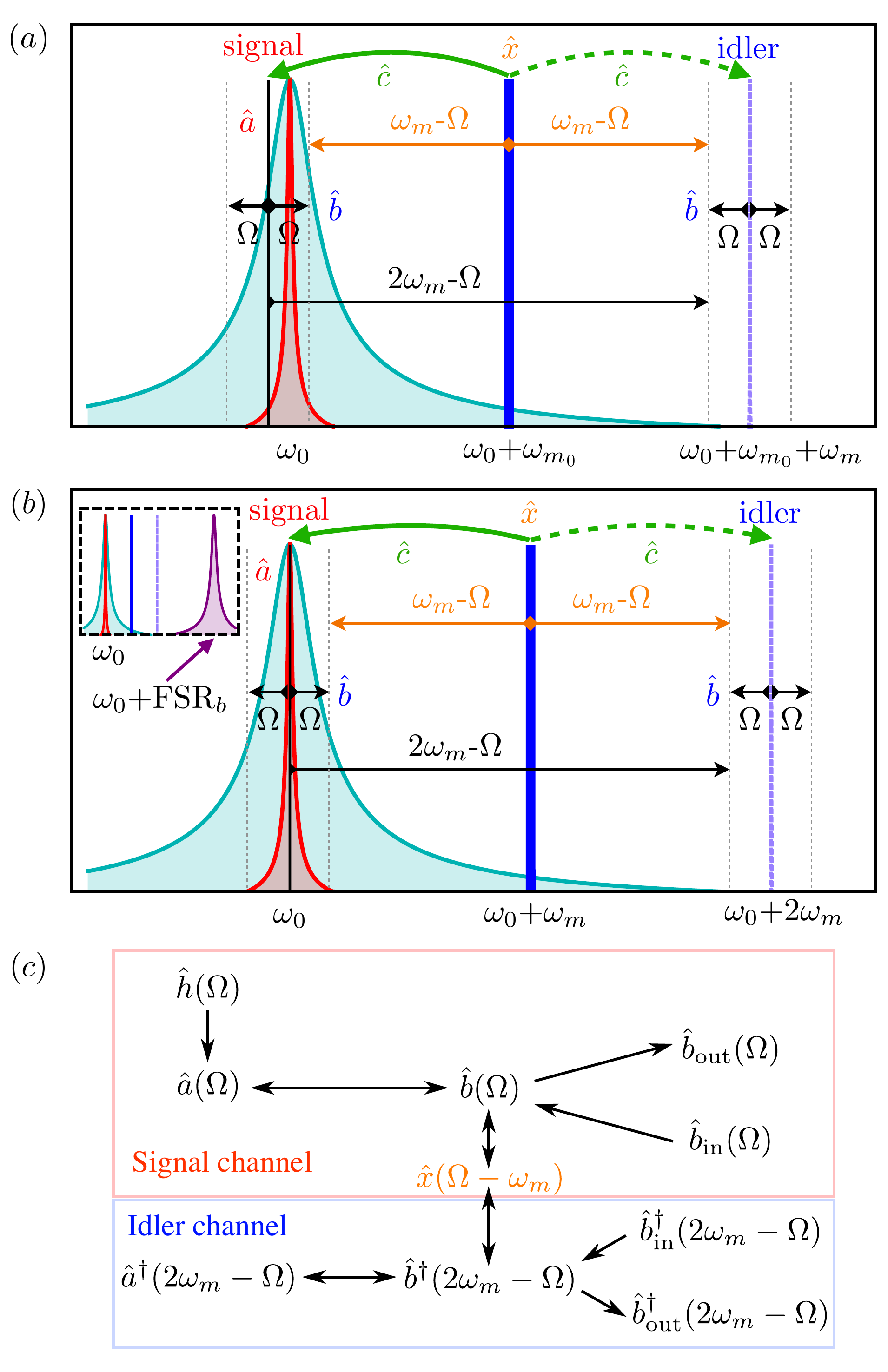}
\caption{Coupling between the optical modes and the mechanical modes represented in frequency domain. %For simplicity, $\hat a$ represents $\hat a_{1, 2}$ and $\hat b$ represents $\hat b_{1, 2, 3, 4}$. 
\textbf{(a)} Without optical spring compensation, the parametric interaction will be shifted by a small frequency (not to scale) which will greatly ruin the designed filter response. \textbf{(b)} With optical spring compensation, the correct sidebands are coupled. It is achieved by making the value of blue-detuning to be slightly larger than the bare mechanical resonant frequency, as described in Eq.\,\eqref{eq:omega_opt}. Insert: the best scenario for resolved-sideband is to set $2\omega_m$ around half $\textrm{FSR}_b$. It will make $\omega_m$ larger than $\gamma_b$ as much as possible and, at the same time, make the influence from the higher FSR as little as possible. \textbf{(c)} The full circulation loop within the signal channel (sidebands around $\omega_0$, represented by the red box in the upper panel) and the idler channel (sidebands around $\omega_0+2\omega_{m}$, represented by the blue box in the lower panel). The idler channel is ignored in the idealised Hamiltonian analysis.}
\label{fig:full_circ_wopt}
\end{figure}

In this section, we will solve the system dynamics in the frequency domain, and analyse the stability using the Nyquist criterion\,\cite{Astrom2010feedback}. 

\subsection{Formalism}
\label{subsec:FD_formalism}

In the frequency domain, Eqs.\,\eqref{eq:field_ops_1}-\eqref{eq:field_ops_2} can be converted into algebra equations in the matrix representation. For the slowly-varying field operators defined in Eq.\,\eqref{eq:omega0frame}, we apply the time-shifting relation in the Fourier transform of optical modes, i.e.
\begin{equation}
{\cal F}[\hat o (t -\tau)]= \hat o(\omega) e^{i\omega \tau}\textrm{ for  }\hat o=\hat a\textrm{ and  }\hat b\,,
\end{equation}
to connect propagating light inside the cavities, where $\omega$ is the sideband frequency in the rotating frame of $\omega_0$. The Fourier transform of the mechanical mode in the filter cavity reads:
\begin{equation}
{\cal F}[e^{-i\omega_{m_0} t}\hat x(t)] = \hat x(\omega-\omega_{m_0})\,.
\end{equation}
With the $\omega_m$ blue-detuned pumping, the optomechanical interaction will couple the optical sidebands at frequency $\omega=\pm\Omega$ ($\Omega\ll\omega_{m_0}$) with the sidebands at  $\omega=2\omega_{m}\mp\Omega$. As mentioned in Fig.\,\ref{fig:config}, the value of $\omega_{m}$ is slightly difference from the mechanical resonant frequency $\omega_{m_0}$. The difference is caused by the optical spring effect,
\begin{equation}\label{eq:omega_opt}
    \omega_m = \omega_{m_0}+\frac{P_b  \omega_0}{2 m \omega_{m_0}^{2} c^{2} \tau_b}.
\end{equation}
when we consider the realistic scenario beyond the resolved-sideband approximation. As illustrated in Fig.\,\ref{fig:full_circ_wopt}\,(a)-(b), when there is optical spring, the parametric interaction will be shifted by a small frequency, causing the demanded sidebands not to be correctly coupled. As our system works near a critical point of PT symmetry\,\cite{Li2020Broadband}, the effect of a rather small optical spring effect is fairly important and needs to be carefully compensated.

As illustrated in Fig.\,\ref{fig:full_circ_wopt}\,(c), the sidebands around $\omega_0$ are named as the signal channel, and the ones around $\omega_0+2\omega_{m}$ are named as the idler channel. We will first treat the filter cavity as an effective mirror, obtain the transformation relation for signal and idler channels, as well as mixing between the channels, and then combine it with the circulation loop of the arm cavity. As the filter cavity is pumped with frequency $\omega_0+\omega_{m}$, for convenience, we temporarily use $\tilde o (\tilde \omega)$ to represent the sidebands in the rotating frame of $\omega_0+\omega_{m}$, where $\tilde o$ can be $\tilde a$ or $\tilde b$, and $\tilde \omega$ is in the order of magnitude of $\omega_{m_0}$. The input-output relation of the filter cavity can be represented as follows: 
\begin{equation}
\left[\begin{array}{c}
\tilde b_{\rm out}(\tilde\omega) \\
\tilde b^{\dag}_{\rm out}(-\tilde\omega)\\
\tilde a_{1}(\tilde\omega) \\
\tilde a^{\dag}_{1}(-\tilde\omega)
\end{array}\right] = \tilde{\bf M}_{\rm filter}(\tilde\omega) \left[\begin{array}{c}
\tilde b_{\rm in}(\tilde\omega) \\
\tilde b^{\dag}_{\rm in}(-\tilde\omega)\\
\tilde a_{2}(\tilde\omega) \\
\tilde a^{\dag}_{2}(-\tilde\omega)
\end{array}\right]\,. 
\end{equation}
where $\tilde{\bf M}_{\rm filter}(\tilde\omega)$ is determined by the optomechanical interaction and the field circulation inside the filter cavity. 

Before connecting the filter cavity to the main interferometer, let's comment on the required functionality of its transfer function in the full circulation loop. Notice that in the lossless case, the open-loop circulation in the arm cavity will result in a phase delay, without harming the amplitude. Thus, we expect the feedback gain provided by the signal channel reflection $\hat a_1(\Omega)\to \hat a_2(\Omega)$ to compensate that effect by providing the same amount of phase advance with an amplitude $1$. For the phase advance, we require ${4 \omega_0 P_{f}  }/({m  \omega_m  c^2 T_{\rm ITM}})=c/L_{\rm arm}$\,\cite{miao2015enhancing} and optical spring compensation to correctly amplify the demanded sidebands. For the physical parameters shown in Table\,\ref{tab:PhyPara}, the optical spring leads to a shift of the mechanical frequency by around 77 Hz. In the actual simulation, this value will be compensated numerically to avoid any tiny discrepancy. The influence of optical spring in the signal-referred shot-noise and stability will be analysed in the following Sections\,\ref{subsec:full_spec},\,\ref{subsec:nyq}. The other requirement is that the amplitude of reflection to be $1$, as we want the idler channel to mix with the signal channel as little as possible. The best scenario for that purpose is to set $2\omega_m$ around half $\textrm{FSR}_b$, as shown by the insert of Fig.\,\ref{fig:full_circ_wopt}\,(b). It will make $\omega_m>>\gamma_b$ as much as possible and, at the same time, suppress the higher FSR from amplifying the idler channel. In addition, we want the mechanical loss to be as small as possible, as the filter cavity contains a parametric process and thus the mechanical loss will ruin the reflection amplitude.
%, the amplitude of reflection will increase with the mechanical loss to maintain the commutation relation.

By setting $\tilde\omega=\mp \omega_m+\Omega$, we can extract relevant matrix components to compose the transformation matrices for the optical fields outside the filter cavity:
\begin{subequations}
\begin{align}
&\left[\begin{array}{c}
\hat b_{\rm out}(\Omega) \\
\hat b^{\dag}_{\rm out}(2\omega_m-\Omega)\\
\hat a_{1}(\Omega) \\
\hat a^{\dag}_{1}(2\omega_m-\Omega)
\end{array}\right] = \tilde{\bf M}_{\rm filter}(-\omega_m+\Omega) \left[\begin{array}{c}
\tilde b_{\rm in}(\Omega) \\
\tilde b^{\dag}_{\rm in}(2\omega_m-\Omega)\\
\tilde a_{2}(\Omega) \\
\tilde a^{\dag}_{2}(2\omega_m-\Omega)
\end{array}\right]\,,\\
&\left[\begin{array}{c}
\hat b_{\rm out}(2\omega_m+\Omega) \\
\hat b^{\dag}_{\rm out}(-\Omega)\\
\hat a_{1}(2\omega_m+\Omega) \\
\hat a^{\dag}_{1}(-\Omega)
\end{array}\right] = \tilde{\bf M}_{\rm filter}(\omega_m+\Omega) \left[\begin{array}{c}
\tilde b_{\rm in}(-\Omega) \\
\tilde b^{\dag}_{\rm in}(2\omega_m-\Omega)\\
\tilde a_{2}(-\Omega) \\
\tilde a^{\dag}_{2}(2\omega_m-\Omega)
\end{array}\right]\,.
\end{align}
\end{subequations}
For simplicity, we define $\tilde{\bf M}_{\mp}\equiv\tilde{\bf M}_{\rm filter}(\mp\omega_m+\Omega)$ and thus have:
\begin{subequations}
\begin{align}
    {\bf R}_{aa}(\Omega)\equiv \left[\begin{array}{cccc}
    \tilde{\bf M}_{-}^{3,3}, &0, &0, &\tilde{\bf M}_{-}^{3,4}\\
    0, &\tilde{\bf M}_{+}^{4,4}, &\tilde{\bf M}_{+}^{4,3}, &0\\
    0, &\tilde{\bf M}_{+}^{3,4}, &\tilde{\bf M}_{+}^{3,3}, &0\\
    \tilde{\bf M}_{-}^{4,3}, &0, &0, &\tilde{\bf M}_{-}^{4,4}
    \end{array}\right]\,,\\
    {\bf T}_{ab}(\Omega)\equiv \left[\begin{array}{cccc}
    \tilde{\bf M}_{-}^{3,1}, &0, &0, &\tilde{\bf M}_{-}^{3,2}\\
    0, &\tilde{\bf M}_{+}^{4,2}, &\tilde{\bf M}_{+}^{4,1}, &0\\
    0, &\tilde{\bf M}_{+}^{3,2}, &\tilde{\bf M}_{+}^{3,1}, &0\\
    \tilde{\bf M}_{-}^{4,1}, &0, &0, &\tilde{\bf M}_{-}^{4,2}
    \end{array}\right]\,,\\
    {\bf R}_{bb}(\Omega)\equiv \left[\begin{array}{cccc}
    \tilde{\bf M}_{-}^{1,1}, &0, &0, &\tilde{\bf M}_{-}^{1,2}\\
    0, &\tilde{\bf M}_{+}^{2,2}, &\tilde{\bf M}_{+}^{2,1}, &0\\
    0, &\tilde{\bf M}_{+}^{1,2}, &\tilde{\bf M}_{+}^{1,1}, &0\\
    \tilde{\bf M}_{-}^{2,1}, &0, &0, &\tilde{\bf M}_{-}^{2,2}
    \end{array}\right]\,,\\
    {\bf T}_{ba}(\Omega)\equiv \left[\begin{array}{cccc}
    \tilde{\bf M}_{-}^{1,3}, &0, &0, &\tilde{\bf M}_{-}^{1,4}\\
    0, &\tilde{\bf M}_{+}^{2,4}, &\tilde{\bf M}_{+}^{2,3}, &0\\
    0, &\tilde{\bf M}_{+}^{1,4}, &\tilde{\bf M}_{+}^{1,3}, &0\\
    \tilde{\bf M}_{-}^{2,3}, &0, &0, &\tilde{\bf M}_{-}^{2,4}
    \end{array}\right]\,,
\end{align}
\end{subequations}
where ${\bf R}_{aa}(\Omega)$ represents the $\hat a_2 \rightarrow \hat a_1$ reflection, ${\bf T}_{ab}(\Omega)$ represents the $\hat b_{\rm in} \rightarrow \hat a_1$ transmission, ${\bf R}_{bb}(\Omega)$ represents the $\hat b_{\rm in} \rightarrow \hat b_{\rm out}$ reflection, ${\bf T}_{ba}(\Omega)$ represents the $\hat a_2 \rightarrow \hat b_{\rm out}$ transmission. Thus, the effect of filter cavity on the optical fields can be expressed as follows:
\begin{subequations}
\begin{align}
\begin{split}
\left[\begin{array}{c}
\hat a_{1}(\Omega) \\
\hat a^{\dag}_{1}(-\Omega)\\
\hat a_{1}(2\omega_m+\Omega) \\
\hat a^{\dag}_{1}(2\omega_m-\Omega)
\end{array}\right] = &{\bf R}_{aa}(\Omega) \left[\begin{array}{c}
\hat a_{2}(\Omega) \\
\hat a^{\dag}_{2}(-\Omega)\\
\hat a_{2}(2\omega_m+\Omega) \\
\hat a^{\dag}_{2}(2\omega_m-\Omega)
\end{array}\right]\\
&+{\bf T}_{ab}(\Omega) \left[\begin{array}{c}
\hat b_{\rm in}(\Omega) \\
\hat b^{\dag}_{\rm in}(-\Omega)\\
\hat b_{\rm in}(2\omega_m+\Omega) \\
\hat b^{\dag}_{\rm in}(2\omega_m-\Omega)
\end{array}\right]\,,
\end{split}\\
\begin{split}
\left[\begin{array}{c}
\hat b_{\rm out}(\Omega) \\
\hat b^{\dag}_{\rm out}(-\Omega)\\
\hat b_{\rm out}(2\omega_m+\Omega) \\
\hat b^{\dag}_{\rm out}(2\omega_m-\Omega)
\end{array}\right] = &{\bf R}_{bb}(\Omega) \left[\begin{array}{c}
\hat b_{\rm in}(\Omega) \\
\hat b^{\dag}_{\rm in}(-\Omega)\\
\hat b_{\rm in}(2\omega_m+\Omega) \\
\hat b^{\dag}_{\rm in}(2\omega_m-\Omega)
\end{array}\right]\\
&+{\bf T}_{ba}(\Omega) \left[\begin{array}{c}
\hat a_{2}(\Omega) \\
\hat a^{\dag}_{2}(-\Omega)\\
\hat a_{2}(2\omega_m+\Omega) \\
\hat a^{\dag}_{2}(2\omega_m-\Omega)
\end{array}\right]\,.
\end{split}
\end{align}
\end{subequations}

%In the idealised analysis discussed in Secion.\,\ref{sec:ideal_to_full}, the propagation phase for small sidebands $\Omega$ was approximated as $e^{i\Omega \tau}\approx 1+i\Omega \tau$, which corresponds to the single-mode approximation $\Omega \tau\ll 1$, that focusing on frequencies lower than the free spectral range of the arm cavity. 

In the idealised analysis in Secion.\,\ref{sec:ideal_to_full}, the idler channel was ignored under resolved-sideband limit. The full analysis here will keep both signal and idler channels. Further linking the filter cavity to the main interferometer, propagation in the arm cavity will be considered for both channels to form a closed-loop transfer function. GW strain signal will be added to the signal channel in the arm cavity and the final input-output relation takes the following form:
\begin{equation}
\left[\begin{array}{c}
\hat b_{\rm out}(\Omega) \\
\hat b^{\dag}_{\rm out}(2\omega_m - \Omega)
\end{array}\right] = {\bf M}(\Omega) \left[\begin{array}{c}
\hat b_{\rm in}(\Omega) \\
\hat b^{\dag}_{\rm in}(2\omega_m - \Omega)
\end{array}\right] + {\bf v}(\Omega)\, h(\Omega)\,, 
\end{equation}
where ${\bf M}(\Omega)$ is a $2\times 2$ matrix describing the transformation of the ingoing field to the outgoing field, and ${\bf v}(\Omega)$ is a $2\times 1$ column vector describing the response of the outgoing field to the GW signal, with the ${\bf v}(\Omega)^{2,1}$ component describing the signal leading into the idler channel $\hat b^{\dag}_{\rm in}(2\omega_m - \Omega)$. The exact expressions for ${\bf M}(\Omega)$ and ${\bf v}(\Omega)$ are quite complicated, and in the subsection that follows, we will use the quantum noise spectral density to illustrate their frequency dependence. 

%In the following sections, for expression simplicity and to avoid confusion, we will use $\omega_{m_0}$ to represent $\Delta$ after the optical spring compensation. 
%\comment{Xiang}{Have utilized $\omega_{m_0}$ to resolve the $\Delta$, $\omega_m$ label issue.}
 
\begin{figure}[!t]
\includegraphics[width=0.9\columnwidth]{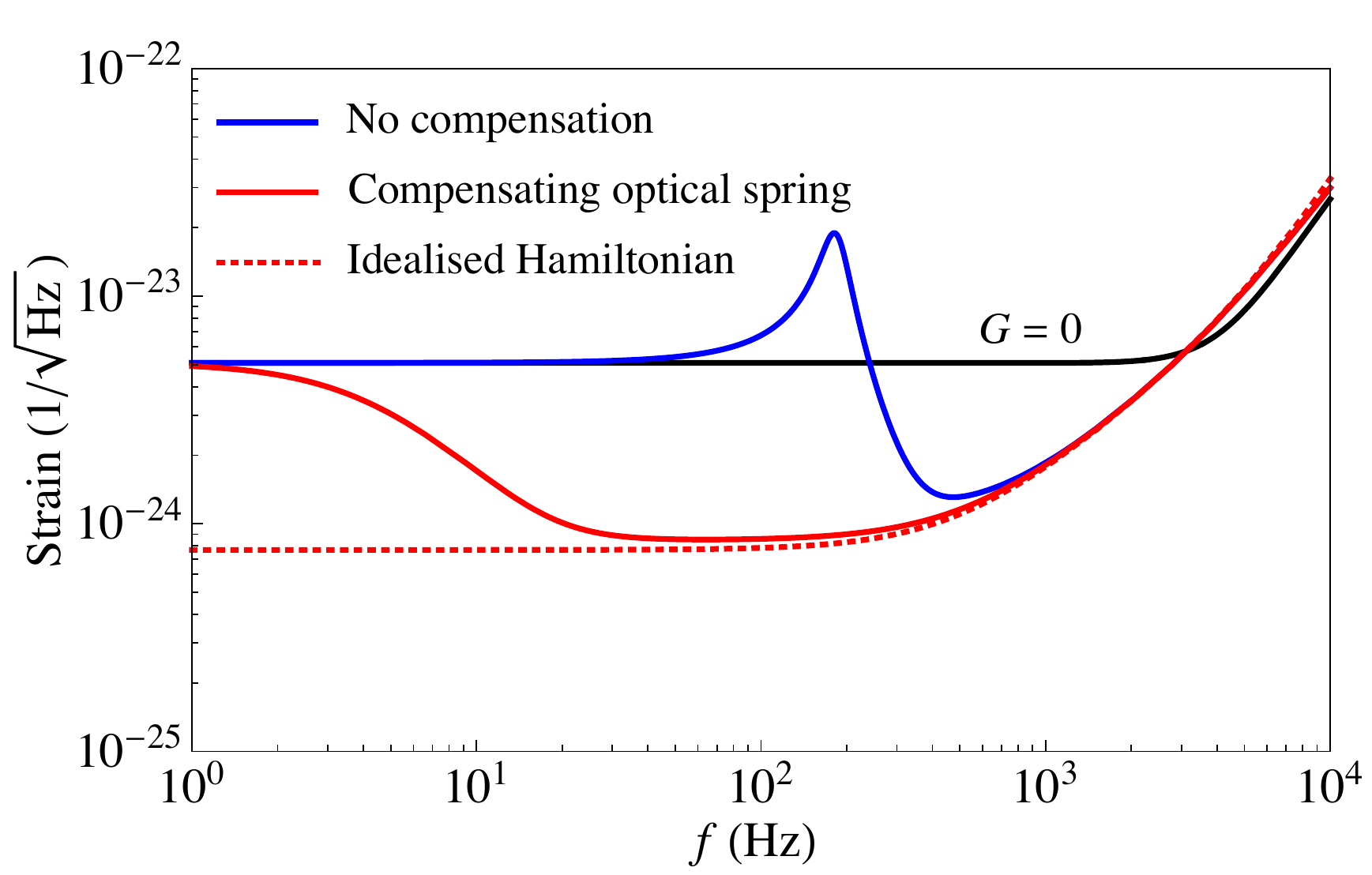}
\caption{The signal-referred shot-noise spectral densities from the full analysis (red and blue) in comparison with the one obtained from idealised Hamiltonian (red dotted curve). Compensating the 
frequency shift of the mechanical oscillator due to the optical spring effect has a significant influence on the sensitivity.}
\label{fig:sens_opt}
\end{figure}

\begin{figure}[!t]
\includegraphics[width=0.9\columnwidth]{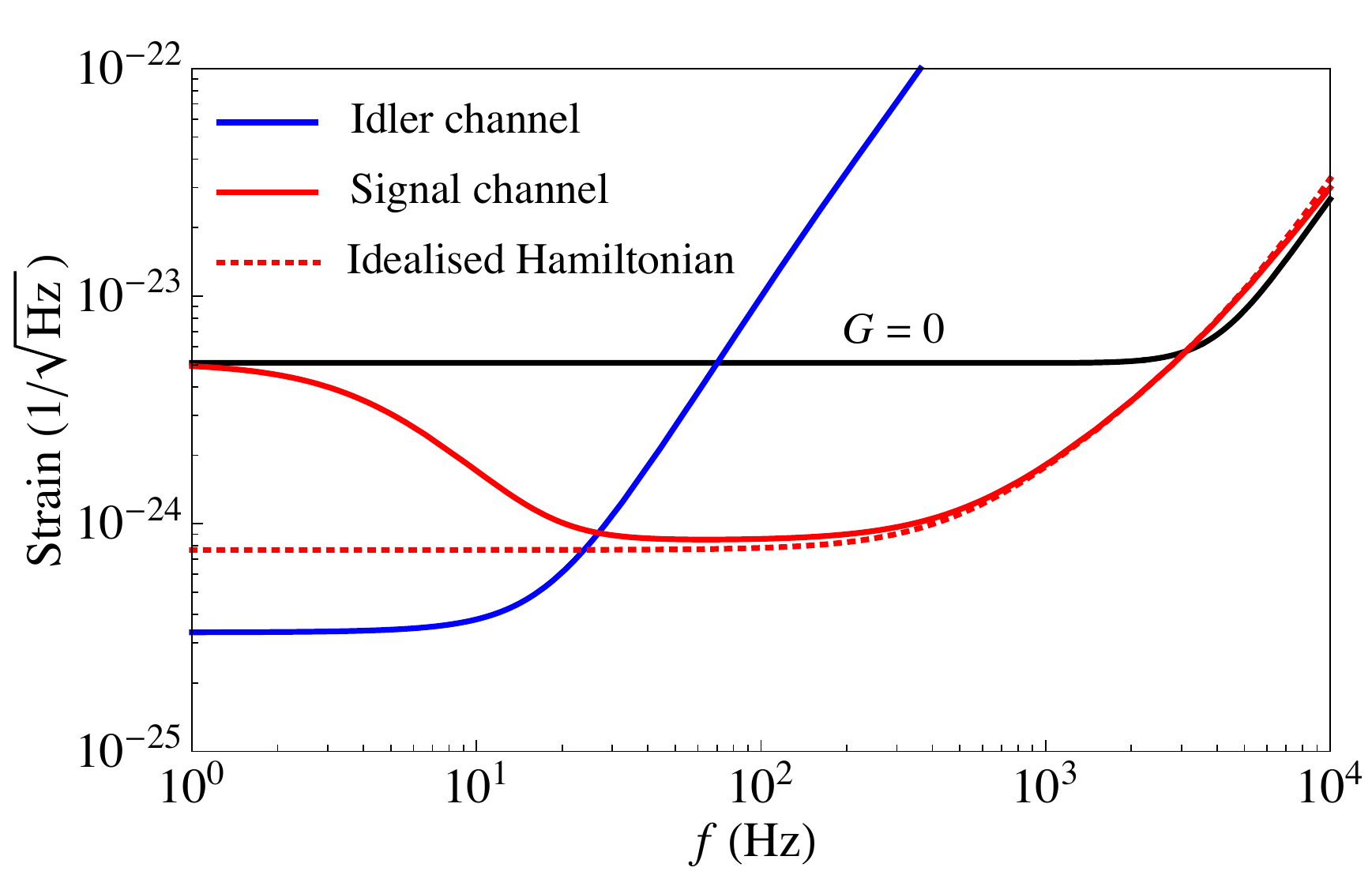}
\caption{The noise spectral density for the idler channel, which can be read out by using a local oscillator at $\omega_0+2\omega_{m_0}$ (blue), in comparison with the signal channel (red). At low frequencies, the signal information contained in the idler channel is even more than that contained in the signal channel.}
\label{fig:sens_idler}
\end{figure}

\subsection{Noise Spectral Density}
\label{subsec:full_spec}

The resulting noise spectral density from the full analysis is shown in Fig.\,\ref{fig:sens_opt} and Fig.\,\ref{fig:sens_idler}. In Fig.\,\ref{fig:sens_opt}, the optical spring effect is shown to play an important role in modifying the sensitivity. As analysed in Section\,\ref{subsec:FD_formalism}, if left unattended, the optical spring will cause an additional resonance in the sensitivity via shifting the central frequency of signal sidebands in the filter cavity response. After compensating such a shift, the full analysis and the idealised Hamiltonian analysis produce a similar result for most of frequencies, given the same set of parameters. However, they start to deviate at low frequencies, where the full analysis shows a higher noise spectrum. 

The deviation from idealised analysis is because of the coupling to the idler channel at low frequencies, as shown in Fig.\,\ref{fig:sens_idler}, where a larger amount of signal will leak to the idler channel than what is output in the signal channel. Thus, the idler channel has a much better sensitivity at low frequencies. We can extract the signal information contained in the idler channel either by using the heterodyne readout with the pump field at $\omega_0+\omega_{m}$ as the local oscillator or using an additional at $\omega_0+2\omega_{m}$ for the homodyne readout of the idler channel. We will discuss the details about the additional homodyne readout scheme in Section\,\ref{sec:opt_readout}, and show the final sensitivity by optimally combining the two readout schemes. 

As we will show in Section\,\ref{subsec:nyq}, the parameters in Table\,\ref{tab:PhyPara} is within the stability regime. Therefore, we can indeed obtain sensitivity improvement without sacrificing the stability.

\subsection{Stability Analysis}
\label{subsec:nyq}

In this section, we use the Nyquist technique\,\cite{Astrom2010feedback} to analyse the system stability. The criterion is a diagrammatic approach to test the stability of a system by only using the open-loop transfer function, even with delay. 

In the idealised analysis in Section\,\ref{sec:ideal_to_full}, $G \approx \,\omega_s$ defines the boundary for stability. Further increasing $G$ through increasing the pumping power of the filter cavity will destabilise the system. In the full analysis when the idler channel is also included, we have a more complicated multi-input-multi-output (MIMO) system. In the sideband picture, given the 2-dimensional open-loop transfer matrix ${\bf M}_{\rm OL}(\Omega)$, the real and imaginary part of the determinant of ${\bf I} + {\bf M}_{\rm OL}(\Omega)$, where $\bf I$ is the 2-dimensional identity matrix, should not enclose the origin of the complex plain. This is limited to the case where ${\bf M}_{\rm OL}(\Omega)$ does not contain elements that have unstable poles, and we therefore also need to be careful about the point where to extract the open-loop transfer function, even though the stability of the final closed-loop transfer function is independent of such a choice. We choose the interface near ITM as highlighted by the red dashed line on the right panel of Fig.\,\ref{fig:fields}, where the top part is the arm cavity and the bottom is consisted of the mechanical oscillator and SRM. 

The open-loop and the close-loop transfer matrices in our case are:
\begin{equation}
{\bf M}_{\rm OL}(\Omega) = {\bf M_{\rm opt}}(\Omega) {\bf M}_{\rm cav}(\Omega), \quad 
{\bf M}_{\rm CL}(\Omega) = ({\bf I} + {\bf M}_{\rm OL}(\Omega))^{-1}\,. 
\end{equation}
The optomechanical transfer matrix ${\bf M}_{\rm opt}(\Omega)$ is defined as:
\begin{equation}
{\bf M}_{\rm opt}(\Omega) =e^{i\Omega \tau_{\rm SRC}/2} \left[\begin{array}{cc} {\cal T}_{\rm opt} & {\cal T}_{\rm opt} + \sqrt{R_{\rm SRM}} \\ -\sqrt{R_{\rm SRM}} - {\cal T}_{\rm opt} & -2 \sqrt{R_{\rm SRM}} - {\cal T}_{\rm opt}\end{array}\right]\,, 
\end{equation}
where
\begin{equation}\label{eq:Topt}
{\cal T}_{\rm opt} = -\sqrt{R_{\rm SRM}}+\frac{ i (1+\sqrt{R_{\rm SRM}})^2 g^2\tau_{\rm SRC}\, \omega_{m_0} }{2[\Omega(\Omega-2\omega_{m})+ i\gamma_m(\Omega-\omega_{m})]}\,.
\end{equation}
Similarly, the passive cavity propagation matrix is:
\begin{equation}
{\bf M}_{\rm cav}(\Omega) =\left[\begin{array}{cc} {\cal T}_{\rm cav}(\Omega) & 0 \\ 0 & 
{\cal T}^*_{\rm cav}(2\omega_{m}-\Omega)\end{array}\right]\,,
\end{equation}
where 
\begin{equation}
{\cal T}_{\rm cav}(\Omega) =  \frac{e^{i\Omega \tau_{\rm arm}}-\sqrt{R_{\rm ITM}}}{1-e^{i\Omega\tau_{\rm arm}} \sqrt{R_{\rm ITM}}}\,. 
\end{equation}

\begin{figure}
\begin{overpic}[width=0.541\columnwidth]
{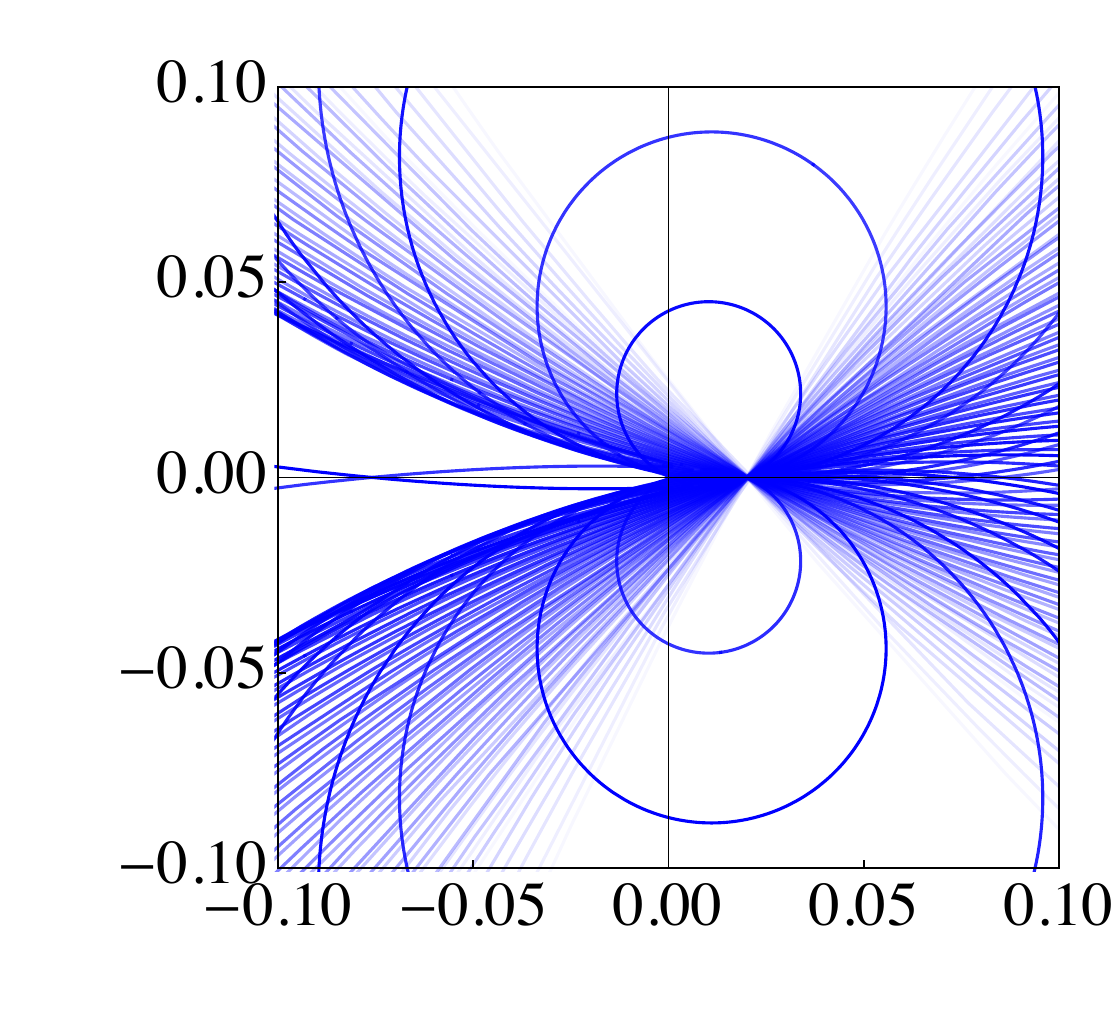}
\put(30,250){\rotatebox{90}{$\Im[\det({\bf I} + {\bf M}_{\rm OL}]$}}
\put(750,5){{$\Re[\det({\bf I} + {\bf M}_{\rm OL}]$}}
\end{overpic}
\includegraphics[width=0.44\columnwidth]{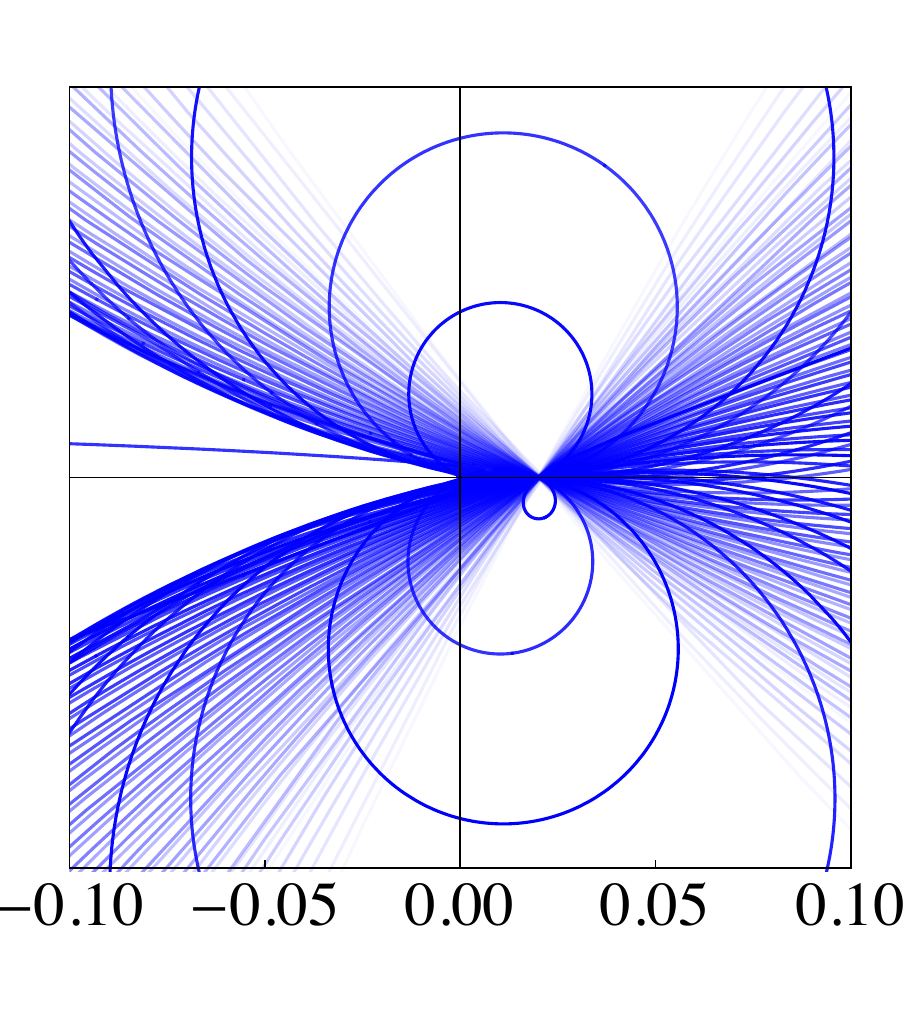}
\caption{The plots for the determinant of the open-loop transfer matrix, given the nominal parameters. The left plot shows the case without compensating the optical spring, while the right one has the optical spring compensated and the contour does not enclose the origin. %The optical spring compensation is necessary for achieving PT-symmetry, and thus meeting the stability condition, because the parametric interaction must be achieved by the correct sidebands and the mechanical modes. 
The opacity of the curve is intentionally made smaller when the magnitude of the frequency is large, which is to highlight the relevant low frequencies.\comment{}{?}}
\label{fig:nyq_os}
\end{figure}

\begin{figure}
\begin{overpic}[width=0.54\columnwidth]
{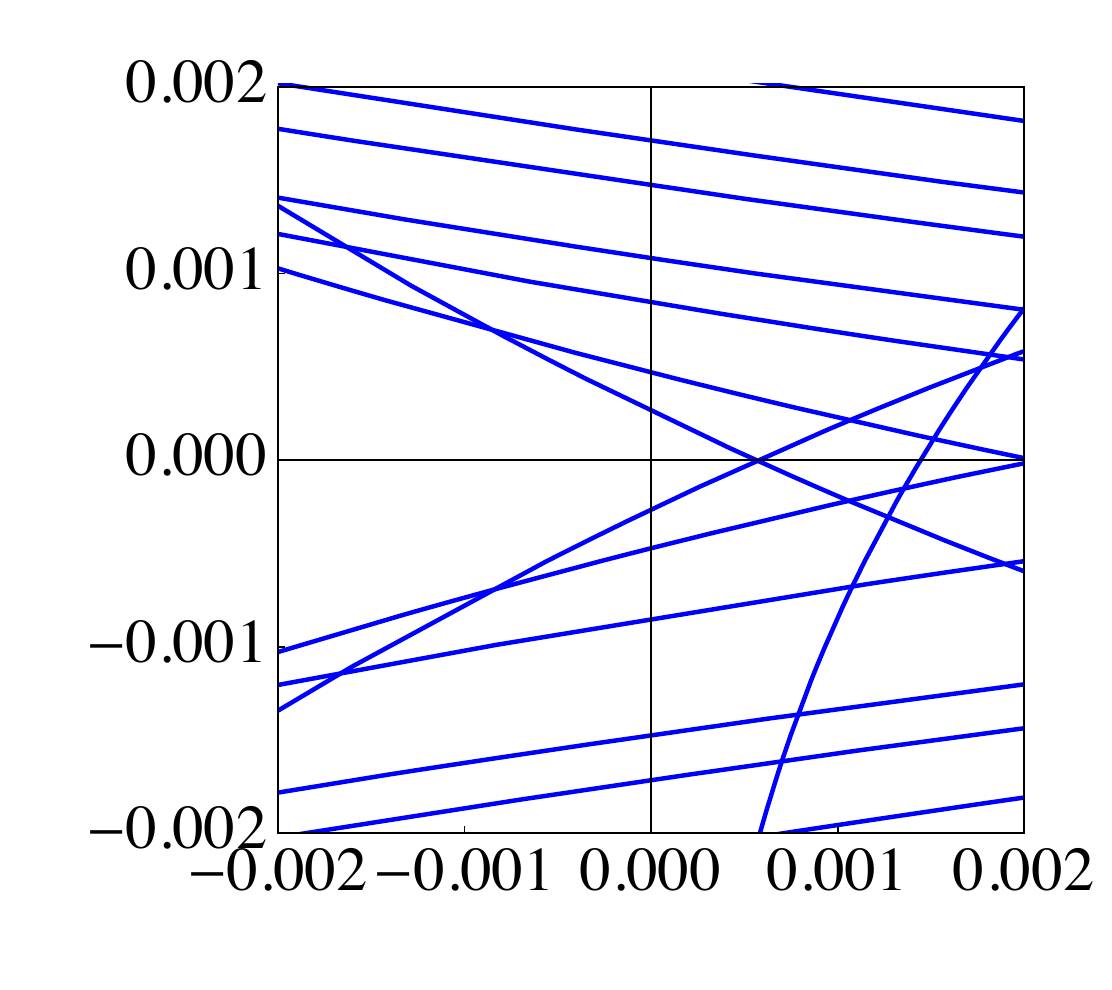}
\put(5,250){\rotatebox{90}{$\Im[\det({\bf I} + {\bf M}_{\rm OL}]$}}
\put(750,5){{$\Re[\det({\bf I} + {\bf M}_{\rm OL}]$}}
\end{overpic}
\includegraphics[width=0.448\columnwidth]{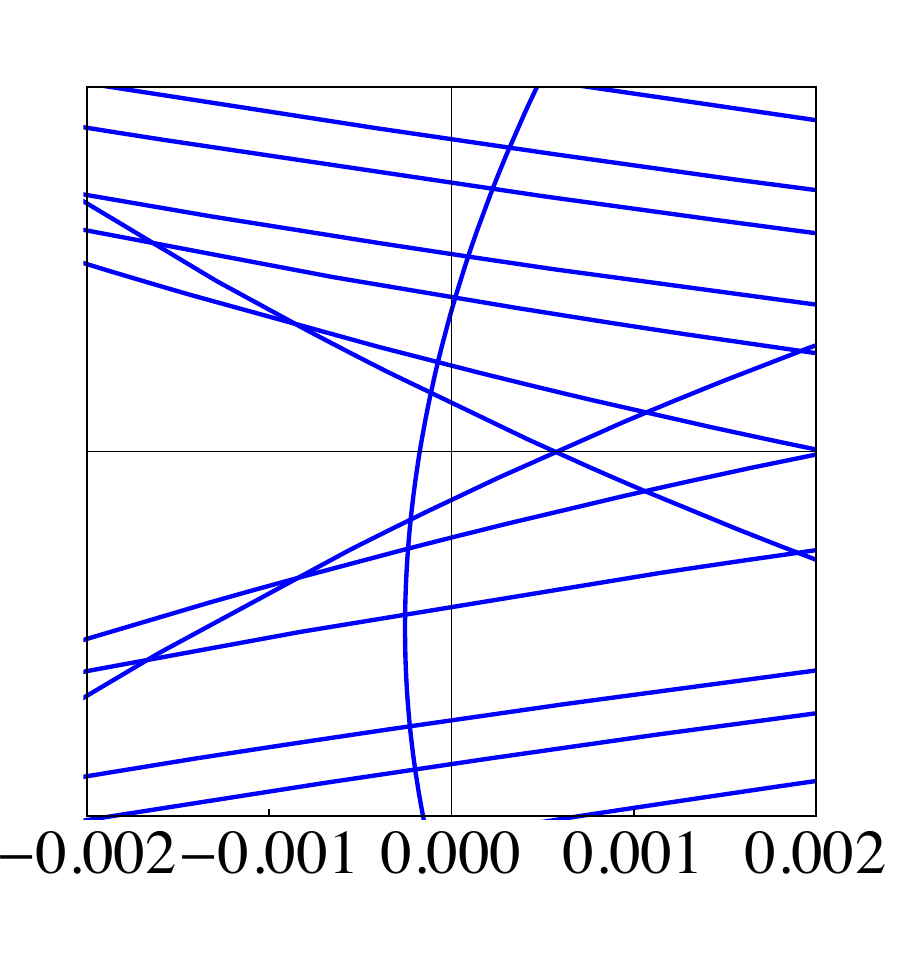}
\caption{The plots for the determinant of the open-loop transfer matrix for $G=\omega_s$ (left) and $G=1.01\,\omega_s$ (right), both with the optical spring shift compensated. For clarity, we only show a tiny regime around the origin, but we have checked that only the right plot has the origin enclosed when zooming out.}
\label{fig:nyq_g}
\end{figure}

In Fig.\,\ref{fig:nyq_os}, we show the resulting Nyquist plot for the nominal parameters in Table\,\ref{tab:PhyPara}. Because the mechanical frequency is not an integer number of the FSR of the arm cavity, the contour is quite complicated due to the phase factor $e^{2i\omega_{m_0} \tau_{\rm arm}}$ of the idler channel. Nevertheless, we only need to check whether the origin is enclosed or not to make a firm claim on the stability, while the complex feature of the contour does not matter. It turns out that, compensating the optical spring shift not only has a significant impact on the sensitivity, as shown in Fig.\,\ref{fig:sens_opt}, but also on the stability. As shown in the plot on the right hand side of Fig.\,\ref{fig:nyq_os}, With the nominal values for the parameters, the system only becomes stable after the optical spring shift is accounted for. This feature can be understood as follows: to meet the PT-symmetry condition, the parametric interaction with the mechanical mode must be achieved by the correct optical sidebands in the filter cavity; thus, only after the optical spring is correctly compensated, it is possible and meaningful to discuss the stability issue.

In Fig.\,\ref{fig:nyq_g}, we further show the effect of the optomechanical coupling rate $G$ on the stability, after the necessary optical spring compensation. Indeed, the system is stable when $G\leq \omega_s$, which is consistent with the idealised Hamiltonian analysis\,\cite{Li2020Broadband}.

\section{Time-domain analysis}
\label{sec:time}

In this section, we perform a numerical time-domain simulation to confirm the enhancement in sensitivity as derived above. The purpose of remaining in the time domain is to capture behaviour that may have been lost due to approximations used in the frequency-domain analysis. It is also much easier for us to capture non-linear behaviour in the time domain, as we have no need to perform Fourier transforms in this analysis. The primary limitation of this method is the relative lack of insight into the physics that we can obtain, as it is difficult to `break open' the simulation and understand the behaviour of individual parts in isolation. As such, our time-domain approach is a powerful complementary tool to the analysis performed thus far and not a replacement.

We will begin by considering Eqs.~\eqref{eq:field_ops_1}-\eqref{eq:field_ops_2}. The principle of this simulation is to appropriately discretise the equations and enable evolve the system forward in time using the knowledge of its previous state. The mechanism through which the system steps forward and the way to recalculate relevant quantities are at the very heart of this simulation and, therefore, it is worthwhile to delve into some details.

\subsection{Discretisation of Equations}
\label{subsec:discr}

The first step towards the time domain simulation is the discretisation of equations. The equations in our system can subdivided into three distinct types. The treatment of the full system will be made clear after the functionality of each type.is explored.

The first type of equation comprises all quantities that are related to others in the way of time-delay. For instance, the field quantities that propagate from one optical component to another. The general form of this type of equations reads:
\begin{equation}
    a_2(t) = a_1(t - \tau),
\label{eq:field_space}
\end{equation}
where $\tau=L/c$ is the time delay caused by the propagation across the space of distance $L$, with $c$ being the speed of light. This type of equations is discretised at time step $n$ as
\begin{equation}
    a_2[n] = a_1[n - n_d],
\label{eq:discret_space}
\end{equation}
where $n_d$ is time-step delay (steps in the past at which the field must be evaluated). It is given by $n_d = \tau / \Delta t$, where $\Delta t$ is the simulation time increment. This type of equations encompasses the forward evolution of the field quantities, whereby a field at a new simulation step can be calculated only with reference to the previous value of another field, which will have been calculated at their step.

\begin{figure}[!t]
\includegraphics[width=\columnwidth/3]{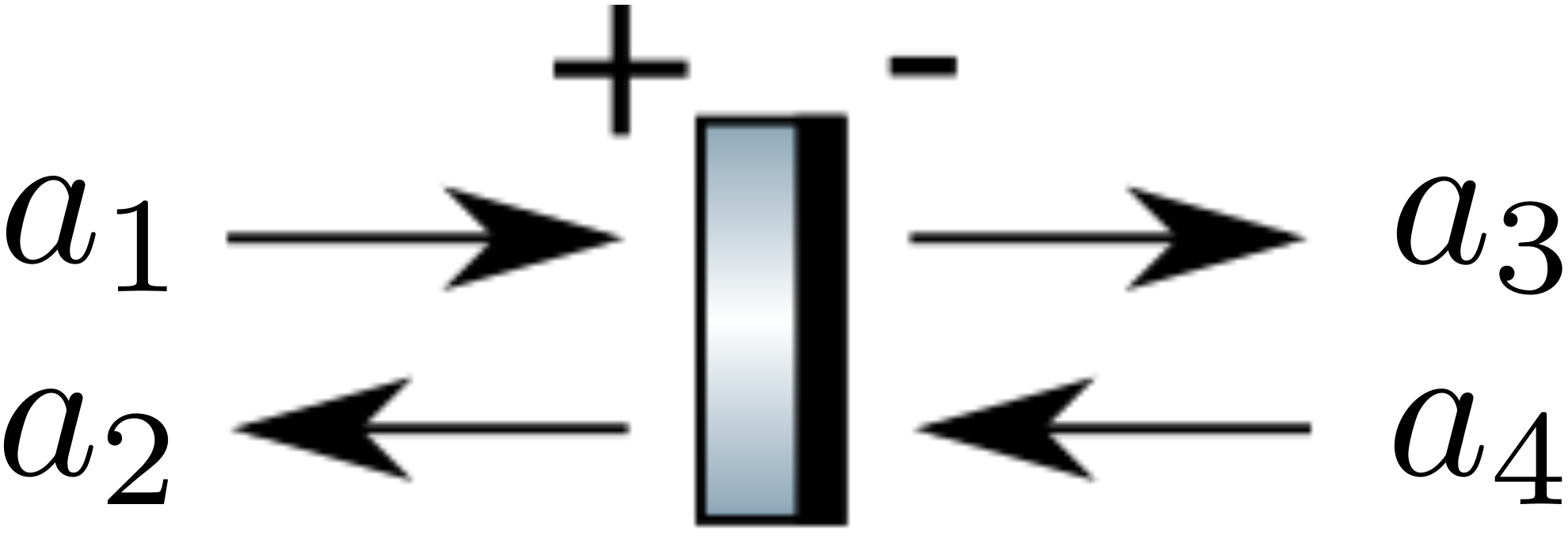}
\caption{Field interactions at a mirror.}
\label{fig:mirror}
\end{figure}

The second type of equation comprises interactions of fields at an optical component. This treatment follows a very standard approach, where the fields at a mirror (as defined in Fig.~\ref{fig:mirror}) are given by:
\begin{subequations}\label{eq:fields_mirror}
\begin{align}
a_2(t) = t_m a_4(t) + r_m a_1(t), \\
a_3(t) = t_m a_1(t) - r_m a_4(t),
\end{align}
\end{subequations}
where $r_m$ and $t_m$ are the mirror amplitude reflectivity and transmissivity respectively. The discretisation of these fields is rather trivial:
\begin{subequations}
\begin{align}
a_2[n] = t_m a_4[n] + r_m a_1[n], \\
a_3[n] = t_m a_1[n] - r_m a_4[n].
\end{align}
\end{subequations}
However, care must be taken in choosing the order to calculate these fields, as the field quantities are all initially unknown. The procedure for calculating these fields will be discussed later in due course.

The final consideration is the treatment of Eq.~\eqref{eq:oscillator_eom}. It can be approached in many ways according to the choice of discretisation for time derivative. Based on its well-studied nature\,\cite{Cieslinski2006}, a ``symmetric'' approach is implemented:
\begin{subequations}
\begin{align}
&\ddot{x}[n] + \gamma_m \dot{x}[n] + \omega_{m_0}^2 x[n] = \frac{F_{\rm rad}[n]}{m},\label{eq:TD_opt} \\
\textrm{with}\ \ &\dot{x}[n] = \frac{x[n + 1] - x[n - 1]}{2 \Delta t},\\
\textrm{and}\ \ &\ddot{x}[n] = \frac{x[n + 1] - 2 x[n] + x[n - 1]}{\Delta t^2}.
\end{align}
\end{subequations}
Note that substituting the latter two equations into Eq.\,\eqref{eq:TD_opt} allows for the $x[n + 1]$ term to be rearranged in terms of quantities at previous time steps and, therefore, it can be used to evolve the position of the mechanical oscillator without further issue.

\subsection{Connecting the System}
\label{subsec:conn}
\begin{figure}[!t]
\includegraphics[width=0.6\columnwidth]{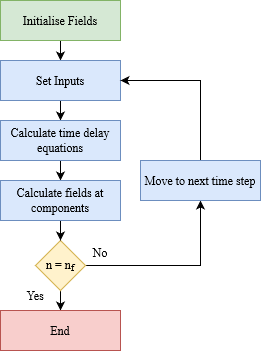}
\caption{Logical flowchart of the time domain simulation.}
\label{fig:flowchart}
\end{figure}

After categorizing the relative equations, the next to consider is connecting the discretized equations into a consistent loop. The most challenging feature of the system is the signal-recycling structure in the optical cavities, as the signal sent ``downstream'' from the simulation inputs returns and combines with the upstream signal in a feedback-like structure. For a system with many cavities, the cycling of signal could become very complicated to handle.

Considering that the signal must traverse the space between optical components, which induces a time delay, it allows us to break up the system into several independent ``compartments''. As the system is evolved forwards in time, when we consider fields at step $n$, we can rely on the values for all steps $<n$ that have already been calculated. This naturally means that all terms of the Eq.~\eqref{eq:field_space} type can be calculated immediately. To tackle the issue with Eqs.~\eqref{eq:fields_mirror}, we can treat the right-hand side quantities $a_{1,4}(t)$ as the ``inputs'' to the mirror, which are already available, as the ``outputs'' or left-hand side of Eq.~\eqref{eq:field_space}. This results in a logic flow, as shown in Fig.~\ref{fig:flowchart}. At the start of a new simulation step, we calculate all the ``outputs'' of the Eq.~\eqref{eq:field_space} type first, which gives all of the ``inputs'' quantities of the Eq.~\eqref{eq:fields_mirror} types. We then calculate the ``outputs'' of all Eq.~\eqref{eq:fields_mirror} types, which then affect the input fields in a neighbouring compartment that links the whole system together. Note that these links occur at a future time step and thus does not need to be calculated now, avoiding the trap of circular logic. This approach greatly simplifies the calculation performed in each simulation cycle. By compartmentalising the equations in this way, we can create several independent systems with very few equations each. It is thus very easy to add extra components to the system, without the need to rewrite any existing elements in the simulation. Furthermore, the system is well set up for optimisation techniques such as parallelisation.

The key issue in our approach is the large number of calculations to be performed, as well as the large number of quantities that need to be stored. To suitably capture the system behaviour, the time increment in simulation needs to be smaller than the shortest time-scale of the system, which in our case is the short cavity traversal time around $100$~ns. Although this MHz level of sampling is far above the frequency range of interest (1 Hz - 10 kHz), it is necessary to capture the behaviour of components that affect the signal response within that frequency band. This will lead to a lot of waste as we are simulating about 1000 times more data than what we are interested in. Furthermore, we want to simulate for at least 1 second, which consists of 10 million samples, resulting in rather a lot of data for a computer to store.

A further limitation is the requirement for integer values of $n_d$ in the discretisation of Eq.~\eqref{eq:discret_space}. For all cavities in the system, the traversal times must be some integer multiple of $\Delta t$. To enable the free choice of cavity lengths, particularly when there are many cavities in the system, $\Delta t$ may have to be even smaller than the shortest cavity traversal time. This is not a big issue in our case as cavity number can be limited to be two. All the optical components are separated by a short delay time $\tau_s$, except for the arm cavity with a long delay time $\tau_a$. For convenience, we can choose $\tau_a$ to be an exact multiple of $\tau_s$, without any loss of physical insight.

\subsection{Step-Response Stability Analysis}
\label{subsec:step}

The steady-state response is the most essentials in the sensitivity analysis of the system. As the simulation loop relies on past values of the system, initialisation is required at some finite time from which the simulation starts. To resolve this issue, all quantities in $t < 0$ are set to be $0$, which corresponds to the off state. After the simulation begins, the building-up process of the steady-state leads to an initial transient effect, which might negatively affect the response estimate. Luckily, in our simulation, this initialisation takes about $0.03$~s, which is sufficiently small compared to the total runtime and will not cause any problem.

\begin{figure}[!t]
\includegraphics[width=\columnwidth]{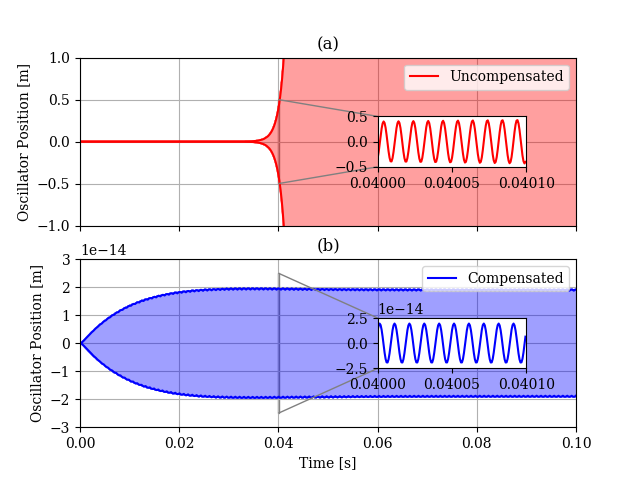}
\caption{In-loop step-response of the position of the undamped mechanical oscillator. \textbf{(a)} When the pump field is offset by $\omega_{m_0}$ only. \textbf{(b)} When the pump field is further offset by compensating the optical spring. The lighter solid fill contains rapid oscillations with the dark bounding lines representing the amplitude envelope. The inserts of (a) and (b) show the zoomed-in views of two sections in the time-series, where the individual oscillations can be seen.}
\label{fig:step_response}
\end{figure}

Interestingly, the transient behaviour behavior is more of an opportunity than a problem. The time-domain response with custom inputs can be used to gain insight into some aspects of the system that can not be easily studied using the frequency-domain approach. One such test we can perform is to determine the stability of the system using its step response. Inserting a constant signal from $t = 0$ (step-function input) to excite the system dynamics, the resulting time-domain behaviour will clearly show whether the system is stable (settling at a constant value or a steady oscillation) or unstable (growing indefinitely towards an infinite amplitude). As shown in Fig.~\ref{fig:step_response}, the position response of the mechanical oscillator in the cases with or without compensating the optical spring clearly shows the stability or instability. It further confirms the result shown in Fig.~\ref{fig:nyq_os} that the system can be stabilised by optical spring compensation.

\subsection{Numerical Noise Spectral Density}
\label{subsec:num}

To check the consistency with the frequency-domain approach in Section\,\ref{subsec:full_spec}, the noise spectral density in this time-domain simulation will be determined in a two-step approach. The noise power spectral density in the strain signal measurement is given by:
\begin{equation}\label{eq:PSD_TD}
    S_{hh}^n(\Omega) = \frac{S_{bb}^n(\Omega)}{T(\Omega) T(\Omega)^*},
\end{equation}
where $S_{bb}^n(\Omega)$ is the noise power spectral density of the appropriate quadrature of $\hat{b}_{\rm out}$ (see Fig.~\ref{fig:fields}) and $T(\Omega)$ is the $h \rightarrow \hat{b}_{\rm out}$ transfer function. The strain noise can thus be obtained through two separate simulations: one to obtain $T(\Omega)$ and another to obtain $S_{bb}^n(\Omega)$.

\begin{figure}[!t]
\includegraphics[width=\columnwidth]{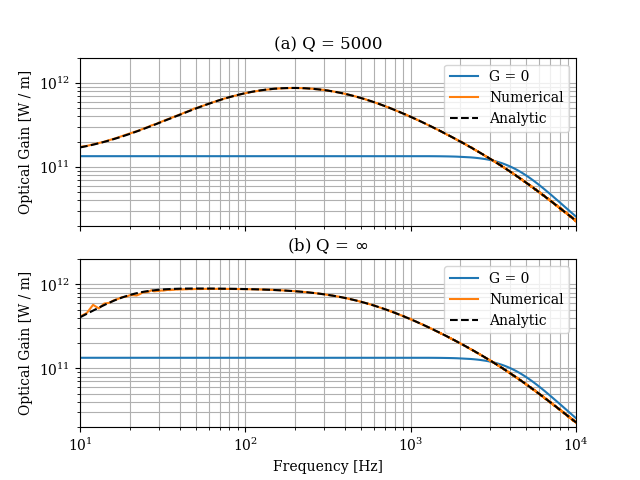}
\caption{Transfer functions for optical gain with optical-spring compensated. The pump-off ($G=0$) gain is shown in contrast to the pump-on gain for \textbf{(a)} a finite mechanical quality factor of 5000, and \textbf{(b)} an undamped oscillator. The numerical, time-domain approach shows very good agreement with the analytical, frequency-domain approach.}
\label{fig:q_comparison}
\end{figure}

\begin{figure}[!t]
\includegraphics[width=\columnwidth]{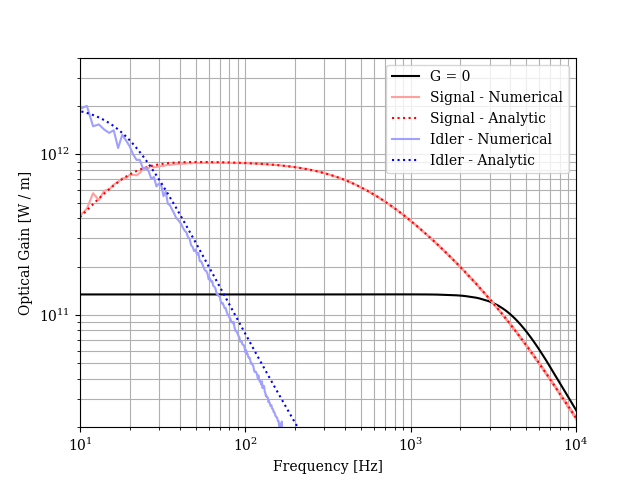}
\caption{Optical gain transfer functions for the signal and idler channels in the undamped, optical-spring-compensated system.}
\label{fig:both_sidebands_optical_gain}
\end{figure}

For a preliminary consistency check, we compare the optical transfer function $T(\Omega)$ by injecting a noiseless random signal with a constant spectral density into the channel of gravitational wave strain. By measuring the spectrum $S_{bh}(\Omega)$ of the output variable $\hat{b}_{\rm out}$, the transfer function $T(\Omega)$ can be determined through the relation between cross- and power spectral density:
\begin{equation}
    S_{bh}(\Omega) = T(\Omega) S_{hh}(\Omega).
\end{equation}
Two simulations are implemented only with a difference in mechanical quality factor, as shown in Fig.~\ref{fig:q_comparison}, and a great agreement has been confirmed between the numerical and analytical approaches. In Fig.~\ref{fig:both_sidebands_optical_gain} we further show the optical transfer function of the mechanically undamped system for both the signal and idler channels, which also shows good agreements.

The second simulation is implemented by injecting all relevant noises with their own spectral densities in the absence of strain signal. This provides $S_{bb}^n(\Omega)$ and, together with the optical transfer function $T(\Omega)$ obtained in the first simulation, allows to complete the calculation of strain noise spectral density according to Eq.\,\eqref{eq:PSD_TD}. The final results of the time-domain simulation are shown in Fig.~\ref{fig:Sideband-noise}, which consistently corresponds to the frequency-domain analysis in Fig.~\ref{fig:sens_idler}. 

\section{Optimal readout scheme}
\label{sec:opt_readout}
%\parts{Amit}

\begin{figure}[t]
\includegraphics[width=0.9\columnwidth]{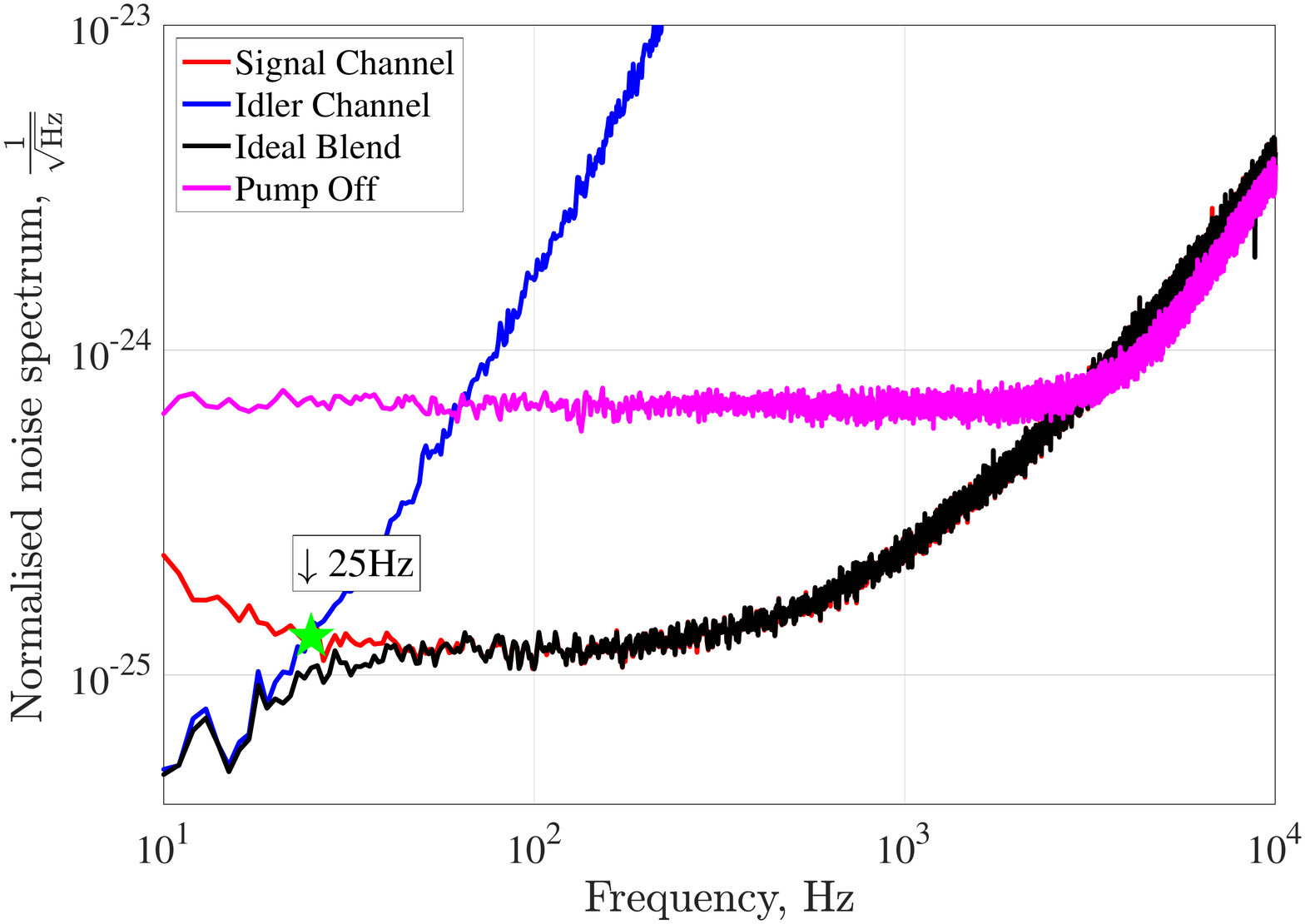}
\caption{The normalized noise spectrum for the signal and idler channels. The signal channel (red) recovers the same response as the pump off case (magenta) which is limited by the bandwidth of the cavity at frequencies of a few kHz and above, which shows there is no degradation of the signal or the bandwidth of the cavity. The idler channel (blue) has greater sensitivity below 30Hz. The ideal blend (black) has greater sensitivity than both the signal and idler channels across the blend frequency range of approximately 25Hz (green star)}.
\label{fig:Sideband-noise}
\end{figure}

In this section, we will investigate the readout scheme to optimally blend the outputs from the signal and idler channels for achieving the best signal sensitivity. Note that the approach we apply here can in general be used for any correlated signals. 

Consider a generic case with two noisy signals which measure the same observable $x$, $\tilde{z_1}$ and $\tilde{z_2}$, such that $\tilde{z_i} = T_i x + \tilde{n_i}$, $i=1,2$, with $T_i$ referring to the transfer function of $x$ to the signal, and $\tilde{n_i}$ being the corresponding noise. These signals can be normalised such that $z_i = \tilde{z_i}/T_i = x + n_i$

In this case $x$ is the gravitational wave signal, where in each measurement there is differing sensitivity in different frequency bands such as that shown in Fig.\,\ref{fig:both_sidebands_optical_gain}. In the frequency domain representation, the normalised signals $z_i$, can be blended together with filters frequency dependent filters, $\alpha(\Omega)$ and $1-\alpha(\Omega)$, which sum to one over the entire frequency range:
\begin{equation}
    Z(\Omega) = \alpha(\Omega)\, z_1(\Omega) + (1-\alpha(\Omega))\,z_2(\Omega),
\label{eq:generic-signal}
\end{equation}
with total noise given by $n(\Omega) = \alpha(\Omega)\, n_1(\Omega) + (1-\alpha(\Omega))\, n_2(\Omega)$. The power spectrum of the total normalised noise can be represented as:
\begin{equation}
\begin{split}
       S_{nn}(\Omega) = &\lvert \alpha(\Omega) \rvert^2 S_{n_1n_1}(\Omega) + \lvert 1-\alpha(\Omega) \rvert^2 S_{n_2n_2}(\Omega) \\
       &+ \alpha(\Omega) (1-\alpha^*(\Omega)) S_{n_1n_2}(\Omega) \\
       &+ (1 - \alpha(\Omega)) \alpha^*(\Omega) S_{n_2n_1}(\Omega),
\end{split}
\label{eq:blending_noise}
\end{equation}
with $S_{n_i n_j}(\Omega)$ being the noise spectrum of ($i=j$) or correlation between ($i\neq j$) $n_1(\Omega)$ and $n_2(\Omega)$. Minimising Eq.\,\eqref{eq:blending_noise} with respect to the conjugate of the filter $\alpha(\Omega)$, i.e. $\frac{\partial S_{nn}(\Omega)}{\partial \alpha^*(\Omega)} = 0$, the optimal filter can be constructed for each frequency $\Omega$ independently:
\begin{equation}
    \alpha(\Omega) = \frac{S_{n_2n_2}(\Omega) - S_{n_2n_1}(\Omega)}{S_{n_1n_1}(\Omega)+S_{n_2n_2}(\Omega)-S_{n_1n_2}(\Omega)-S_{n_2n_1}(\Omega)}.
\label{eq:optimal_filter}
\end{equation}
%As indicated by Eq.\,\eqref{eq:optimal_filter}, the filters and the power spectra are frequency dependent. %This will be assumed for the rest of this description of the optimal combination of the signals.
Applying the filters to Eq\,\eqref{eq:generic-signal}, we can obtain the the power spectrum of the total signal $Z(\Omega)$:
\begin{equation}
\begin{split}
        S_{ZZ}(\Omega) = &\lvert \alpha(\Omega) \rvert^2 S_{z_1z_1}(\Omega) + \lvert 1-\alpha(\Omega) \rvert^2 S_{z_2z_2}(\Omega) \\
        &+ \alpha(\Omega) (1-\alpha^*(\Omega)) S_{z_1z_2}(\Omega) \\
        &+ (1 - \alpha(\Omega)) \alpha^*(\Omega) S_{z_2z_1}(\Omega),
\end{split}
\label{eq:total-Power-Spectrum}
\end{equation}
whose noise power spectrum is the same as $S_{nn}(\Omega)$ in Eq.\,\eqref{eq:blending_noise}.
When there is correlation between $n_1$ and $n_2$, the noise across the applied high and low pass filters $\alpha(\Omega)$ and $1-\alpha(\Omega)$ can be further improved. When the cross-correlation terms are zero, the final noise spectrum would be determined by the ability to create the optimal blends between two filters, such as high order roll off of the filters with phase margin at the blending frequency. The advantage of this signal post-processing scheme is to reduce the stability requirement if the system is operating in a closed feedback loop.

In the simulation discussed in Section\,\ref{sec:time}, the signal can be extracted from both the signal channel around $\omega_0$ and the idler channel near $\omega_0 + 2\omega_{m_0}$. Applying the optimal blending strategy of Eq.\,\eqref{eq:optimal_filter}, the overall noise spectrum can take advantage of whichever channel with the better behaviour at one frequency. As shown in Fig.\,\ref{fig:Sideband-noise}, the idler channel contributes more below 25Hz, while the signal channel dominates at higher frequencies above 25Hz, and the optimal blending curve follows the better one at all frequencies.

\section{Conclusions}
\label{sec:discussion}

In this paper, we analyse in detail the realization of a PT-symmetric interferometer\,\cite{Li2020Broadband} using an optomechanical filter cavity. We go beyond the idealised Hamiltonian analysis under single-mode and the resolved-sideband approximations, and consider the real physical parameters in the filter cavity setup. We prove that, after compensating the optical spring in the filter cavity, the stability and sensitivity improvement stated in the original proposal\,\cite{Li2020Broadband} remains valid in the realistic settings, even when a portion of the signal gets mixed into the idler channel.

To perform full analysis, we implement numerical simulations in both frequency-domain and time-domain. The methods consistencies have been checked with both the signal transfer function and output noise spectrum. In the frequency-domain analysis, the system stability is confirmed using Nyquist criteria, where the Nyquist plot with the stable parameter setting does not enclose the origin. %of the complex plane of the determinant of the open-loop transfer matrix. 
In the time-domain simulation, the stability manifests itself in the transient behaviour before the system reaches its steady-state, where the stable parameter setting won't make any system quantity go to infinity.

Considering the unavoidable leakage of the signal into the idler channel in the actual setups, we further constructed the blending scheme for the output of the two channels. Applying the optimal output filter, one can take advantage of the channel with better behaviour at each frequency, to obtain optimal sensitivity at all frequencies. Another scheme with an auxiliary mechanical cooling beam is under investigation, where the information in the mechanical oscillator can be read out supplemently. Together with the idler channel, they can provide more possibilities for optimized blending readout.

%\comment{}{mentioning prospective work?}

\section*{Acknowledgements}
We would like to thank Chunnong Zhao for helpful comments on the manuscripts during LIGO P\&P review. X.L. and Y.C.'s research is funded by the Simons Foundation (Award Number 568762), and the National Science Foundation, through Grants PHY-2011961, PHY-2011968, and PHY-1836809. J.S., A. S. U., J. B., H. M., and D. M. acknowledge support from the Birmingham Institute for Gravitational Wave Astronomy and UK EPSRC New Horizons award (Grant No. EP/V048872/1). H.M. has also been supported by UK STFC Ernest Rutherford Fellowship (Grant No. ST/M005844/11). Y.M. has been supported by HUST university start up funding.

\bibliography{PTWLC.bib}

\end{document}